\documentclass[10pt]{article}
\usepackage{amsfonts}
\usepackage{amssymb}
\usepackage{amsthm}
\usepackage{amsmath}
\usepackage{mathabx} 
\usepackage{color}

\usepackage[pdftex]{graphicx}

\newcommand{\sss}{\setcounter{equation}{0}}
\newtheorem{theorem}{THEOREM}[section]

\newtheorem{lemma}[theorem]{LEMMA}

\newtheorem{corollary}[theorem]{COROLLARY}

\newtheorem{remark}[theorem]{REMARK}

\newtheorem{prop}[theorem]{PROPOSITION}

\newtheorem{definition}[theorem]{DEFINITION}

\def\hv{\hat{\mathbf v}}

\def\hw{\hat{\mathbf w}}

\newcommand{\dom }{{\rm dom}}

\newcommand{\ere}{ {\mathbb R}}

\newcommand{\ese}{{\mathbb S}}

\def\p2{\mathcal A_{\Phi}(B)}
\def\0p2{\mathcal A_{\Phi}(0)}
\def\sp2{\mathcal A_{\Phi}(B)}
\def\beq{\begin{equation}}
\def\ene{\end{equation}}

\newcommand{\bull}{\hfill $\Box$}
\def\qed{\ifhmode\unskip\nobreak\fi\ifmmode\ifinner
\else\hskip5pt\fi\fi\hbox{\hskip5pt\vrule width4pt height6pt
depth1.5pt\hskip1pt}}

\def\hv{\hat{\mathbf v}}

\def\hw{\hat{\mathbf w}}

\def\mo{\mathbf p}

\def\+out{x^{\rm out}}

\begin{document}

\baselineskip=20 pt
\parskip 6 pt

\title{ {  High-Momenta Estimates for the Klein-Gordon Equation: Long-Range Magnetic Potentials and Time-Dependent Inverse Scattering  }
\thanks{ PACS Classification (2008): 03.65Nk, 03.65.Ca, 03.65.Db, 03.65. AMS Classification (2010): 81U40, 35P25
35Q40, 35R30.  Research partially supported by the project PAPIIT-DGAPA UNAM IN102215 }}
 \author{ Miguel Ballesteros and Ricardo Weder  \thanks{Fellows of the Sistema Nacional de Investigadores.}   \\
 Departamento de F\'{\i}sica Matem\'atica. \\
 Instituto de Investigaciones en Matem\'aticas Aplicadas y en Sistemas.\\
  Universidad Nacional Aut\'onoma de M\'exico. Apartado Postal 20-126\\ IIMAS-UNAM, Col. San Angel, C.P. 01000, M\'exico D.F., M\'exico\\
miguel.ballesteros@iimas.unam.mx,  weder@unam.mx}

\maketitle

\begin{center}
\begin{minipage}{5.75in}
\centerline{{\bf Abstract}}
The study of obstacle scattering for the Klein-Gordon equation in the presence of long-range magnetic potentials is addressed. Previous results of the authors are extended to the long-range case and the results the authors previously proved for high-momenta long-range scattering for  the Schr\"odinger equation are
 brought to the relativistic scenario. It is shown that there are important differences between relativistic and non-relativistic scattering concerning long-range. In particular, it is proved that the electric potential can be recovered without assuming the knowledge of the long-range part of the magnetic potential, which has to be supposed in the non-relativistic case. 
The electric potential and the magnetic field are recovered from the high momenta limit of the scattering operator, as well as fluxes modulo $2 \pi $ around handles of the obstacle. Moreover, it is  proved that, for every $\hv \in \mathbb{S}^2$,
$ A_\infty(\hv) + A_\infty(-\hv)$ can be reconstructed, where $A_\infty$ is the long-range part of the magnetic potential. A 
a simple formula for the high momenta limit of the scattering operator 
is given, in terms of magnetic fluxes over handles of the obstacle and long-range magnetic fluxes at infinity, that are introduced in this paper. The appearance of these long-range magnetic fluxes is a new effect in scattering theory.

\bigskip

\end{minipage}
\end{center}

\section{Introduction}  \sss

\subsection{The Aharonov-Bohm Effect : Essential Features}\label{int1}

The Aharonov-Bohm effect is a fundamental issue in physics. It describes a phenomenon that is not compatible with classical mechanics, but that 
 can be predicted from quantum physics. Moreover,  it describes the physically significant (classical) electromagnetic quantities in quantum mechanics. It is, thus, not only a specific phenomenon but a highly influential concept in quantum theory, and its experimental verification is an important  confirmation of its accuracy as a physical theory. We now describe in more detail what we just mentioned. 
  
The situation we analyse is an electron (a test charged particle) in the presence of a classical magnetic field (although in this paper we consider also electric fields, now take for the moment only into account a magnetic field). According to classical mechanics (Newton's law) the motion of the particle is totally determined by the force (acting in the particle) and the initial position and velocity. In this case the force is given by the Lorentz formula:
\begin{equation}\label{in1}
q {\bf v}\times B,
\end{equation}
where ${\bf v}$ is the velocity of the particle, $ q $ is the charge and $  B $ is the magnetic field (evaluated at the position of the particle). This gives, as classical mechanics predicts, the complete picture of the electron motion. Therefore, the only physical quantity affecting the behaviour of the particle is the magnetic field.   
However, the world is not so simple and what classical mechanics predicts is not correct. There are other physically significant quantities besides the force. These quantities are not easy to precise, but let's discuss a little bit more about it (in the text below we consider how to determine them): 

In quantum physics the situation we just described can be modeled through the Schr\"odinger equation 
\begin{equation}\label{in2} 
i \frac{\partial}{\partial t} \,\phi = \big (  \mo - A \big )^2 \phi,
\end{equation}
where $ \mo = -i \nabla$ is the momentum operator, $A$ is a magnetic potential such that $\nabla \times A = B $. Here we set $\hbar $ to one, the mass of the particle equal $1/2$,  and include in $A$ the electric charge. After a close look at the Schr\"odinger equation one might think that the physical significant quantity we are looking for is the magnetic potential. This, nevertheless, is not correct. The reason is that in quantum mechanics the wave function $ \phi$  is not uniquely determined, but depends on a representation, i.e., it is defined up to a unitary transformation, like a gauge transformation. Then, changing $A$ by $A + \nabla \lambda$, for some scalar function $\lambda$,  must not change any physical prediction. Thus, describing physically significant quantities is subtle business. Here is where the contribution of Aharonov  and Bohm
\cite{ab} takes place  (see also \cite{f}): They consider an infinitely long straight thin solenoid and a magnetic field confined to it. They describe a situation in which an electron wave packet consisting in two separated beams is directed to the solenoid. Each beam passes through different sides of the solenoid and they are brought together behind the solenoid in order to produce an interference pattern. It turns out that the interference pattern depends on the  magnetic field  enclosed in the solenoid, even if the electron never touches it. Let's use 
loosely the Schr\"odinger equation to explain this issue (for a rigorous justification see \cite{bw3}). Let us suppose that the solenoid is located at the vertical axis and that the whole situation does not depend on the vertical variable. Then, we reduce the problem to two dimensions.  Since the magnetic field vanishes outside the solenoid, the magnetic potential is gauge-equivalent to zero in every connected region on its exterior, but not in the full exterior of the solenoid. 

When the magnetic field inside the solenoid is zero, the solution to the Schr\"odinger equation  \eqref{in2} is supposed to consist of two beams,
$$
\phi_0(t,x)= \phi_{0,1}(t,x)+ \phi_{0,2}(t,x),
$$
where $\phi_{0,j}, j=1,2$ are separately solutions to the Schr\"odinger equation and they remain in connected regions away from the solenoid. Furthermore, as time increases,  $\phi_{0,1}$  is  supposed to pass through the left of the solenoid and $\phi_{0,2}$ to the right of it. Aharonov and Bohm argued \cite{ab} that when the magnetic field inside the solenoid is not zero the solution to the Schr\"odinger equation is given again by two beams,
$$
\phi(t,x)= \phi_1(t,x)+ \phi_2(t,x), \,\hbox{\textrm where}\, \phi_j(t,x)= e^{-i \lambda_j(x)}\, \phi_{0,j}(t,x), j=1,2,
$$
and $ \lambda_j$ is the circulation $ \int A\cdot dx$ of the magnetic potential along the path of the beam $\phi_{0,j}, j=1,2$.
Furthermore, assuming that initially (at $t=0$) both beams are close to each other and located far from the solenoid near a point $ x_{-\infty}$ we can take the functions $\lambda_j$ as follows,
\begin{equation}\label{in7}
\lambda_j(x) : =  \int_{C_j(x_{- \infty}, x)} A\cdot dx, 
\end{equation}
where $ C_j(x_{- \infty}, x)  $ is a simple differentiable path joining the points $x_{- \infty} $ and $x$ , where for $j=1,2$  the path goes, respectively,  to the left and to the right of the solenoid. 
The two beams are brought together  at some  point, $x_\infty$,  behind the solenoid, where  
$\phi_1$ acquires  a phase $ e^{- i  \int_{C_1(x_{- \infty}, x_\infty)} A\cdot dx }  $ and $\phi_2$ acquires  a phase $ e^{- i  \int_{C_2(x_{- \infty}, x_\infty)} A\cdot dx }  $.

We denote by $\mathcal{C}$ the simple closed curve obtained joining     $  C_1(x_{- \infty}, x_\infty)  $ and  $  C_2(x_{ -\infty}, x_{\infty}) $ with counterclockwise orientation.
Then, the difference in face between the left and the right beams at the point $x_\infty$ is given by (we use Stokes' theorem)
\begin{equation}\label{in8}
e^{- i \int_{\mathcal{C}} A}   = e^{ -  i\Phi_B }, 
\end{equation}    
where $ \Phi_B $ is the magnetic flux in a transverse section of  solenoid. The approximate solution described above is the prediction of Aharonov and Bohm. In \cite{bw3} we called it the Ansatz of Aharonov and Bohm. Note that  it consists of multiplying the free solution (when the magnetic field inside the solenoid is zero) by the Dirac magnetic factor \cite{Dirac}. In \cite{bw3} we proved rigorously that the Aharonov-Bohm Ansatz is indeed and approximate solution and we gave error bounds. Actually, in \cite{bw3} we considered the case of a toroidal magnet, as in the  works where the Aharonov-Bohm prediction  was experimentally verified  
\cite{Tonomura1,Tonomura2,Tonomura3,Tonomura4}. 

As pointed out by Aharonov y Bohm, the phase factor \eqref{in8} can be predicted from quantum mechanics, even though the particles never touch the solenoid, i.e., the force produced by the magnetic field at the position of the particle is zero, all the time. This can be interpreted   in the following two ways: 
\begin{enumerate}
\item The magnetic field acts non locally.
\item Some properties of the magnetic potential are physically significant.  
\end{enumerate}
Regardless which interpretation we choose, it is important to precise the quantities determining the physics of the problem. Apparently, they are the electromagnetic fields and the fluxes of the magnetic potential, modulo $2\pi$, over closed paths. This is in agreement with the discussion above, the complete description of electrodynamics in terms of non-integrable factors (see \cite{wuyang} and \cite{Dirac}) and the experimental results in \cite{Tonomura1,Tonomura2,Tonomura3,Tonomura4}.

\subsubsection{Relevance of  High Velocity and Relativistic Scattering }

In the explanation above we assumed that we can control the fate of the beams $\phi_1$ and $\phi_2 $ and that they stay in a connected region not touching the solenoid all the time. This is not possible to achieve, but only approximately. However, the accuracy of the approximation depends on how ballistic is the motion of  each beam, in order to control the spreading of the beams. Then, have to assume high enough momenta. The magnitude  of the  momenta, that we have to choose in order to have a good approximation, depends on the particular geometry and physical parameters of the system. This is why a relativistic theory (allowing relatively high momenta) is relevant.           

\subsection{Historical Context and the Necessity of Toridal Geometries} 

There is very a large literature on the Aharonov-Bohm effect. We, of course, do not pretend to be exhaustive, but to report the main advances in relation to our work. For an extensive review up to 1989 see \cite{olariu-popescu} and \cite{peshkin-tonomura}. We give here more recent references, but only the key contribution regarding our work. The readers are advised to look at the literature of our references, if they are searching for a complete record.  \\

The two dimensional model of Aharonov and Bohm (see \cite{ab}) has the disadvantage of requiring infinite straight solenoids (see the discussion in Section \ref{int1}). They, of course, do not exist in nature.  The argument that sufficiently long (straight) solenoids could be considered infinite is controversial for the following reason: The topology of the
exterior of a finite solenoid is trivial in the sense that all closed curves can be continuously deformed to a point. This implies, from Stoke's theorem, that the magnetic field cannot be confined in the solenoid and, therefore, the field leakage produces a flux that equals the magnetic flux inside the solenoid. Moreover, scattering experiments regularly send and detect particles from faraway of the target. In the description of the Aharonov-Bohm prediction we did above (Section \ref{int1}) we used a wave packed separated in two beams. The beams are sent and detected faraway from the solenoid. Even though the magnetic field is very weak,  the magnetic flux outside the magnet enclosed by the trajectory of both beams could be of the same order than the flux enclosed in the magnet, since the beams travel long distances. Then,       magnetic flux enclosed by the paths the electrons follow could be significantly different from the flux in the solenoid itself. This is a situation when long enough might not signify infinite, because the topology of the exterior of long enough solenoids and the topology of the exterior of an infinite solenoid are dramatically different.  Another reason why the infinite solenoid scenario might be problematic (and this is what we analyse and prove here and in \cite{bw-2dim}) is that in two dimensions the magnetic potentials must be long-range and the long-range potentials influence scattering (the scattering operator) in a way that some information non related to fields or magnetic fluxes modulo $2\pi$ can be inferred from the scattering operator (this is already present in \cite{ab}). We believe that this information might be non-physical as we explained in the previous section. 

The amount of papers, books and experiments dealing  with the case of a straight solenoid (as proposed in \cite{ab})
is large. However, scientists recognized the problem of the field leakage already decades ago. Since then, the issue became controversial and a new geometry proposal emerged: The toroidal geometry. This geometry allows to confine a magnetic field without leakage (notice that the topology of the exterior of a toroidal magnet is not trivial in the sense that there are closed curves that cannot be continuously deformed to a point). This led to the seminal experiments, with toroidal magnets,  carried out by Tonomura et al. \cite{Tonomura1, Tonomura2, Tonomura3, Tonomura4}. In these remarkable experiments they split a coherent electron wave packet into two parts. One travelled
inside the hole of the magnet and the other outside the magnet. They brought both parts together behind the magnet
and they measured the phase shift produced by the magnetic 
flux enclosed in it, giving a strong evidence
of the existence of the Aharonov-Bohm effect. 

The experiments of Tonomura et al. \cite{Tonomura1, Tonomura2, Tonomura3, Tonomura4} reduced the 
controversy  to a lower scale. The interpretation of the results was the new trend for some scientists. Some works proposed an interpretation in which the results by Tonomura et al. could be explained by the action of a force. See, for example, \cite{bo, he} and the references quoted there. The force they referred to
 would accelerate the electron producing a time delay. In a recent experiment Caprez et al.
\cite{caprez} obtained that there is no acceleration. Then, they proved experimentally that the explanation of the results of the Tonomura et al. experiments by the action of a force is wrong.  
  
From the theoretical point of view some efforts have been directed to
justify that the Aharonov-Bohm Ansatz approximates correctly  the solution to the Schr\"odinger equation. There have been numerous works trying to provide such approximations. Several Ans\"atze
have been proposed, without giving error
bound estimates. Most of these attempts are qualitative, although some of them give numerical values.
Fraunh\"ofer diffraction, first-order Born and high-energy approximations, Feynman path integrals and
the Kirchhoff method in optics were used. For a review of the literature up to 1989 see  \cite{olariu-popescu} and \cite{peshkin-tonomura}. Recently we rigorously proved that the Ansatz of Aharonov
and Bohm is a good a approximation to the solution of the Schr\"odinger equation and we analysed the full scattering picture, see \cite{bw, bw2, bw3}.  In particular,  in \cite{bw2} we gave a rigorous quantitave proof, under the experimental conditions,  that quantum mechanics predicts the experimental results of Tonomura et al. \cite{Tonomura1, Tonomura2, Tonomura3, Tonomura4}. In this work and in  \cite{bw-kg} we address the relativistic case. This paper is dedicated to long-range potentials, whereas \cite{bw-kg} considers short-range potentials.       

\subsubsection{Relevance of  Long-Range Potentials }
In the proposal of Aharonov and Bohm (see \cite{ab}), the infinite straight solenoids restrict the space to two dimensions (see Section \ref{int1}). However, the two dimensional situation requires long-range magnetic potentials. Through these potentials the scattering operator encodes information that is not related to fluxes modulo $2\pi$ nor to the electromagnetic field. If this was a physical information, we would have new phenomena.  However, we believe that this information is not physical, see Section \ref{int1}. The question at stake is: Which physical information can be extracted from the scattering operator ? i.e. to what extent can we rely on the scattering operator ?. Notice that the Aharonov-Bohm effect gives an example in which the scattering cross section is not the only information we can extract from scattering. The analysis of this in two dimensions for the non-relativistic case is done in \cite{bw-2dim}. Here we address the question for the relativistic scenario.  Notice that there are important differences between both cases, as we explain in this text.

\subsection{The Role of Long-Range Magnetic Potentials -- Description of Our Model and Further Historical Context}

We study obstacle scattering of charged relativistic particles in the presence of long-range magnetic potentials. The obstacle $K$ is assumed to be a finite union of handle bodies, for example of tori and balls. Inside it there is an inaccessible magnetic field.   
In particular, we focus on the effects of the long-range part of the magnetic potentials in high-momenta scattering. This article extends the results in \cite{bw-kg}, where only short-range magnetic potentials are addressed, and proves in the relativistic case results similar to the ones   in \cite{bw-2dim}, that considers the Schr\"odinder equation. We prove that all results for the Klein-Gordon equation in  \cite{bw-kg} are valid in the long-range case, but furthermore, we demonstrate that some information from the long-range part of the magnetic potentials can be reconstructed from high-momenta scattering. The role of long-range magnetic potentials in inverse-scattering has lately acquired interest (see \cite{bw-2dim}, \cite{EI2}, \cite{EI1} and \cite{EI3}). The question at stake is: What are the properties of the magnetic potentials that can be recovered from the scattering operator ? This is a subtle question because the magnetic potentials are not physically significant in classical physics. Therefore, the above question is of purely quantum mechanical nature. Moreover, according to the complete description of electromagnetism in terms of non-integrable phase factors introduced in \cite{wuyang} (see also \cite{Dirac}) and the experimental results on the Aharonov-Bohm effect (see \cite{caprez}, and \cite{Tonomura1}-\cite{Tonomura4}),  the only observable quantities (related to the magnetic potential in our setting) are magnetic fluxes modulo $ 2 \pi$ over the handles of the obstacle. Nevertheless, it is proved in \cite{bw-2dim} and  
\cite{EI2}-\cite{EI1} that the long-range part of the magnetic potential (that is not related to the magnetic field or magnetic fluxes around handles) can be recovered  from the (non-relativistic) scattering operator, in certain situations. In  this paper we go further and prove similar results for relativistic equations, more precisely the Klein-Gordon equation. This brings new insights to the understanding of long-range magnetic effects in quantum mechanics, because differences and similarities, with respect to the non-relativistic case, appear. For example, it is shown in \cite{bw-2dim} that in the high-velocity limit of the (non-relativistic) scattering operator  the long-range part of the magnetic potential and the electric potential are coupled,
which implies that we have to assume the knowledge of the long-range part of the magnetic potential in order to be able to recover the electric potential and the other way around (we have to assume the knowledge of the electric potential in order to recover the long-range part of the magnetic potential). This also happens in \cite{EI2}-\cite{EI1} (see the discussion about this fact in the introduction of \cite{bw-2dim}). Fortunately, this problem seems to be artificial because considering special relativity in our equations (using the Klein-Gordon equation) decouples the electric and the magnetic potentials in the high-momenta limit. Then, the high momenta limit of the scattering operator permits the reconstruction of the  electric potential  without requiring the knowledge of the magnetic potential, which is one of the results in this paper. 
 Additionally, in contrast to the non-relativistic case, in the relativistic situation, the fluxes around the handles of the obstacle can be recovered modulo $2 \pi $ only if the electric potential (and the magnetic field) vanish, otherwise we can only recover them modulo $\pi$. This is physically reasonable, because of the relativistic  duality between the electric and the magnetic fields.  Recovering the magnetic field from high momenta scattering does not distinguish between relativistic and non-relativistic models, as we show in this text.  Denoting by $A_\infty(\hv)$, $\hv \in \mathbb{S}^2$, the long-range part of the magnetic potential (see Proposition \ref{rem1}), relativistic scattering only allows us recovering $A_{\infty}(\hv) + A_\infty(-\hv) $ and not $ A_\infty(\hv) $, $\hv \in \mathbb{S}^2$, as is the case for the non-relativistic Schr\"odinger equation. 
 
 Finally, we give a simple formula for the high momenta limit of the scattering operator in terms of magnetic fluxes over handles of the obstacle and long-range magnetic fluxes at infinity, that we introduce in this paper. The appearance of these long-range magnetic fluxes is a new effect in scattering theory.	 This result is also true in the non-relativistic case of the Schr\"odinger equation in three dimensions considered in \cite{bw}, with a similar proof, and also in the two dimensional case studied in \cite{bw-2dim}, with the necessary changes due to the  differences in the geometry.

The Aharonov-Bohm effect  \cite{ab}, \cite{f} for non-relativistic equations and short-range magnetic potentials is studied, for example, in \cite{bw}, \cite{bw2}, \cite{bw3}, \cite{EI2}, \cite{EI1}, \cite{EI3}, \cite{n} and \cite{w1}, and the references quoted there.  
For high-momenta scattering for relativistic equations in the whole space, see \cite{jung1} and \cite{jung2}. The magnetic Schr\"odinger equation, in the whole space,  is studied in \cite{arians}. The time dependent methods for inverse scattering that we use are introduced in \cite{ew}, for the Schr\"odinger equation. A survey about many different applications of this time dependent method for inverse scattering can be found in \cite{w4}. The direct scattering problem for the Klein-Gordon equation is studied in \cite{Ger}, \cite{w2} and \cite{w3} and the references cited there.

\subsection{Main Results and Description of the Paper}\label{mr}

Here we describe our main results and give a short guideline of the paper. The obstacle is properly defined in Section \ref{obstacle}. It consists in a union of handle bodies. In Figure 1 we draw an example of the class of obstacles we consider. For each handle of $K$ (let's say the handle number $j$), we choose a curve $\hat \gamma_{j}$ surrounding it.  The classes of magnetic potentials we use depend on the magnetic fluxes over the curves $ \hat \gamma_{j}$, $j \in \{ 1, \cdots , m\}$. We define
in Section \ref{ass-2.1} three classes of magnetic potentials (see Definitions \ref{def-2.3} and \ref{hvlwso-d.1}): The most general one is denoted by $\mathcal{A}_\Phi^{{\rm LR}}(B)$ (the superscript LR stands for long-range), 
  $\mathcal{A}_\Phi^{{\rm SR}}(B)$ denotes the class of short-range magnetic potentials  and, finally, we define a class of regular long-range magnetic potentials  $\mathcal{A}_{\Phi, \delta}(B)$. Here $B$, defined in the same section, is the magnetic field outside the obstacle and $\Phi$ represents the circulations of the magnetic potentials over the curves
 $ \hat \gamma_{j}$, $j \in \{ 1, \cdots , m\}$. We denote by $A_0$ the electric potential, it is defined in Section \ref{ass-2.1}. As in \cite{bw-kg}, \cite{w2}, \cite{w3} we write the free and the interacting Klein-Gordon equations as first order in time  $2$ by $2$ systems, respectively, in the free Hilbert space $\mathcal{H}_0$ and in the interacting Hilbert space $\mathcal{H}(\underline{A})$.   See Section \ref{hamilt}. The free Hamiltonian, $H_0$, is introduced in \eqref{ham0}. The perturbed Hamiltonian, $H(\underline A)$, is defined in
 \eqref{ham}. Here $\underline A = (A_0, A)$, $A$ being the magnetic potential.
In Section \ref{wavescat} we prove the existence of the wave and scattering operators.
The wave operators are the defined by the strong limits 
\beq
  W_{\pm}(\underline A)= \hbox{\rm s-}\lim_{t \rightarrow \pm \infty} e^{it
H(\underline A)}\, J\, e^{-it H_0},
\label{2.19prima}
\ene  
where $J$ is a bounded  identification  operator from  $\mathcal{H}_0$  into  $\mathcal{H}(\underline{A})$. The scattering operator is 
$$S(\underline A) = W_+(\underline A)^* W_-(\underline A). 
$$  
One of our main results is Theorem \ref{TPG} in which we prove (with error bounds) that the high momenta limit of the scattering operator, in the representation where the free Klein-Gorgon operator is diagonal (see \eqref{diagu}), is given by 
\begin{equation}\label{extra}
\begin{pmatrix}
 e^{i \int_{-\infty}^{\infty} dr  (   A \cdot \nu - A_0) ({\bf x} + r \nu ) }  & 0 
\\ 0 & e^{- i \int_{-\infty}^{\infty} dr   ( A \cdot \nu + A_0) ({\bf x} + r \nu ) }
\end{pmatrix}, 
\end{equation}
where $ {\bf x}  $ is the multiplication operator with respect to the variable $x \in \mathbb{R}^3$. 
which extends Theorem 2.8 in \cite{bw-kg} to the long-range case. Eq. \eqref{extra} is used in Section \ref{smfepf} to recover the electric potential and the magnetic field in Theorem \ref{inverse-fields} and  magnetic fluxes over the handles of the obstacle, modulo $2\pi$,  in Theorem \ref{th-7.1}. Our main results on long-range effects in high-momenta scattering are presented in Section \ref{rmLR}, in particular in Theorems \ref{TPhiA} and \ref{th-7.12l}.  For every magnetic potential 
$A \in\mathcal{A}_{\Phi, \delta}(B) $, $\delta > 1$, we define its long-range part by
$$
A_\infty (\hv) = \lim_{s \to \infty } sA_\infty (s\hv).      
$$  
In Theorem \ref{TPhiA} we prove that $A_\infty(\hv) + A_\infty(- \hv)$, for every $\hv \in \mathbb{S}^2$, can be recovered from the high momenta limit of the scattering operator.  In Theorem \ref{th-7.12l} we give a simple formula for the high momenta limit of the scattering operator, assuming the electromagnetic field vanishes,
in terms of magnetic fluxes modulo $2 \pi$ around the handles of the obstacle and a long-range magnetic fluxes at infinity that we introduce in Definition \ref{long-defi}. As mentioned above,  this result is also true in the non-relativistic case of the Schr\"odinger equation in three dimensions  considered in \cite{bw}, with a similar proof, and also in the two dimensional case studied in \cite{bw-2dim}, with the necessary changes due to the  differences in the geometry.

\section{ Model}  \label{dmmr}\sss

\subsection{Description of the Model}\label{SDM}
We extend the model introduced in \cite{bw-kg} to the study of long-range magnetic potentials. In this section, several definitions and notations are borrowed from \cite{bw-kg}. We briefly repeat some of them, for the convenience of the reader, and include the new ingredient: The long-range magnetic potentials. We refer  the reader to \cite{bw-kg} for a more detailed presentation. 

\subsection{General Notation} 

For every normed vector space $X$ we denote by $\| \cdot \|_{X}$ its norm. If no confusion arises, we omit the subscript in case $X = L^2(O)$ or $L^2(O) \oplus L^2(O)  $, for some open subset $O$ in $\mathbb{R}^3$, or $X$ is a space of operators. We assume the same convention for inner products.  In this text we denote by $B(x; r)$ the open ball in $\mathbb{R}^3$ centred at $x$ and with radius $r$.  For any vector ${\mathbf v} \in \mathbb R^3\setminus \{0\}$ we designate,  $\hat{\mathbf v}= \frac{\mathbf v}{|\mathbf v|}$.  For any set $O \subset {\mathbb R}^3$ we denote by $O^c$ its complement, by $O^o$ its interior,  and by $\chi_O$ the characteristic function of $O$.
By  $C$  we denote a positive, non-specified, constant.      We use the standard notation $\langle x \rangle = (1 + x^2)^{1/2} $, for every $x \in \mathbb{R}^3$. 
The Schwartz space of rapidly decreasing $C^\infty $-functions in $\mathbb{R}^3$ is denoted by 
$\mathcal{S}(\mathbb{R}^3)$.
For every $n \in \mathbb{N}$ and every open set $ O $ in $\mathbb{R}^3$, we denote by 
${\bf H}^n(O)$ 
the Sobolev  space of functions with distributional derivatives up to order $n$ square integrable, and by 
${\bf H}_0^n(O)$
the closure of $C_0^\infty(O)$ in ${\bf H}^n(O)$, see \cite{adams}. For every strictly  positive function $\omega : O \mapsto \mathbb{R} $ we denote by $ {\bf H}^n_{\omega}(O) $ the corresponding weighted Sobolev space. For every 
$\phi \in {\bf H}^n_{\omega}(O) $, 
$$
\| \phi \|_{{\bf H}^n_{\omega}(O)} := \sum_{\alpha_1 + \alpha_2 + \alpha_2 \leq n} \Big \|  \omega^{1/2} \frac{\partial^{\alpha_1} }{\partial x_1^{\alpha_1} }\frac{\partial^{\alpha_2} }{\partial x_2^{\alpha_2} }\frac{\partial^{\alpha_3} }{\partial x_3^{\alpha_3} } \phi  \Big \|_{L^2(O)}. 
$$

 We analyze charged relativistic particles moving outside a compact obstacle, $K$, in three
dimensions. We denote by
$\Lambda := \ere^3 \setminus K$. Inside $K$ there is
an inaccessible magnetic field and in $\Lambda$ there is an electromagnetic field. 
We denote by $A_0$  the electric potential and by $B$ the magnetic field.

We recall the notation we choose for the momentum operator and the position operator. The momentum operator is denoted by 
$$
\mo = -i \nabla
$$
and the position operator is denoted by ${\bf x}$, which is the multiplication operator by the variable $ x \in \mathbb{R}^3 $. The
momentum operator is, clearly, the multiplication operator by the variable $ p \in \mathbb{R}^3$, in momentum space representation.
We utilize along the  paper 
${\bf v}$ to denote a fixed vector in $\mathbb{R}^3$ and 
$\hv = \frac{{\bf v}}{\|  v \| } $ (if ${\bf v} \ne 0  $). We use 
$\nu $ to denote a fixed element of the sphere $\mathbb{S}^2$ and 
$ v $ will be positive real number, generally the norm of $ {\bf v}$. We do not adopt the convention that a bold face symbol always represent a vector.

\subsubsection{The Obstacle $K$}\label{obstacle}
The obstacle $K$ is a compact submanifold of $\ere^3$.
We denote by $ \{ K_j \}_{j=1}^L $ its connected components. 
 We assume that the $K_j$'s are
handle bodies. For more details  see \cite{bw}, where the same obstacle is
addressed. See Figure 1.
\subsubsection{The Electromagnetic Field and the Potentials}

\begin{definition}[The Magnetic Field]\label{ass-2.1}
{ \rm The magnetic field,
 $B$, is a real-valued, closed and bounded $2-$ form defined in $ \overline{\Lambda}$.
We assume that it is two times continuously differentiable. We additionally suppose that
\beq
\int_{\partial K_{j}} B=0, \, j \in \{1,2,\cdots,L\},
\label{2.1}
\ene
and
\beq
\Big| \Big ( \frac{\partial }{\partial x_1} \Big )^{a} 
\Big (  \frac{\partial }{\partial x_2} \Big )^{b}
\Big (  \frac{\partial }{\partial x_3} \Big )^{c}  B(x)   \Big|  \leq C (1+|x|)^{- \mu}, \, \hbox{\rm for some}\,\, \mu > 2, \: \: \text{and every $a, b, c \in \{0, 1, 2 \} $ 
with $a + b + c \leq 2$}.
\label{2.2}
\ene
}
\end{definition}
\begin{definition}[Electric Potential] {\rm
The electric  potential is a real-valued function, $A_0$, satisfying 
 (for some $\varepsilon > 0$)
\beq \label{pos}
\langle A_0^2\phi  ,  \phi \rangle \leq \langle  -\Delta \phi,  \phi    \rangle 
+ ( m^2 - \varepsilon ) \langle   \phi,  \phi    \rangle, 
\ene
for every $\phi \in {\bf H}_0^1(\Lambda)$ .We, furthermore, assume that for some $ \delta <  1/5 $ there exists a constant $C_\delta$ such that  
\beq \label{sub}
\langle A_0^2\phi  ,  \phi \rangle \leq  \delta \langle  -\Delta \phi,  \phi    \rangle 
+ C_\delta \langle   \phi,  \phi    \rangle, 
\ene
for every $\phi \in {\bf H}_0^1(\Lambda)$.
 We assume,  additionally, that for some $C^{\infty}$
function $\kappa$, defined in $\mathbb{R}^3$, such that $ \kappa=0$ in a neighborhood of $K$ and   with $1 - \kappa$ compactly supported, $\kappa A_0$ is two times continuously  differentiable and  
\beq
\Big| \Big ( \frac{\partial }{\partial x_1} \Big )^{a} 
\Big (  \frac{\partial }{\partial x_2} \Big )^{b} \Big (  \frac{\partial }{\partial x_3} \Big )^{c}  \kappa A_0(x)   \Big|  \leq C (1+|x|)^{- \zeta}, \, \hbox{\rm for some}\,\, \zeta > 1, \: \: \text{and every $a, b, c \in \{0, 1, 2 \} $ 
with $a + b + c\leq 2$}.
\label{2.4}
\ene }
\end{definition}
For sufficient conditions in order that \eqref{pos} and \eqref{sub}  hold see \cite{chech}, \cite{w1}, \cite{w2}.

\subsection{Classes of Magnetic Potentials} \label{mp}

We adopt the definitions, and notations,  presented in Eq. (2.6) in \cite{bw} : 
We set  $\{ \hat{\gamma}_j\}_{j=1}^L$ the  curves described in Figure 1. 

\begin{definition} \label{def-2.2}{\rm
We denote by  $\Phi$ a function $\Phi: \{\hat{\gamma}_j\}_{j=1}^L \rightarrow \ere$. 
We call it the flux.}
\end{definition}

\begin{definition} \label{def-2.3}{\rm
We denote by $\mathcal{A}_{\Phi}^{{\rm LR}}(B)$ the class of continuous $1-$ forms, $A$, defined in $ \overline{\Lambda}$ such that  $d A = B$, $  \int_{\hat{\gamma_j}}\, A= \Phi (\hat{\gamma_j}), \hspace{.2cm} \forall   \, j \in \{1,2,\cdots, L \}, $ and 
\beq
|A(x)| \leq C (1+|x|)^{- 1}, \hspace{2cm}  a(r) : = \sup_{x \in \overline \Lambda, |x| \geq r} A(x)\cdot \frac{x}{|x|} \in L^1(\mathbb{R}).
\label{2.10b}
\ene
 If additionally 
\beq \label{noq}
A(x) \leq C(1 + |x|)^{-\zeta}, 
\ene
for some $\zeta > 1$, we say that $A \in \mathcal{A}_{\Phi}^{{\rm SR}}(B) $.
Here the superscript LR stands for long-range and the superscript SR stands for 
short-range.
} 
\end{definition}
\begin{remark}\label{coupot} {\rm In  Theorem 3.7 of \cite{bw} (see also Remark 2.4 of \cite{bw-kg}) the Coulomb potential is constructed. This potential belongs to $ \mathcal{A}_{\Phi}^{{\rm SR}}(B) $ for some $\zeta >1$ that depends on $\mu$.
 
}
\end{remark}

\begin{definition}\label{Diotaab}
{ \rm
For every $a, b \in [0, \infty)$ with $ a  + b > 2 $, we define the function
$\iota_{a,b}: \mathbb{R}^3 \to \mathbb{R}$: 
\begin{align}\label{iotaab}
\iota_{a,b}(x) : = \begin{cases}  \frac{1}{(1 + |x|)^{\min(a, b)}} + \frac{1}{(1 + |x|)^{a + b - 2}}  , & \text{if}\:\: a,b \neq 2,  \\
\frac{1}{(1 + |x|)^{2}} + \frac{\ln(e + |x|)}{(1 + |x|)^{a + b - 2}}, &  \text{
if  $ \: a= 2$ or $\: b = 2$}.  \end{cases}  
\end{align}   }
\end{definition}
\begin{definition}[Second Class of Long-Range Magnetic Potentials]\label{hvlwso-d.1}{\rm
For every vector potential $ A \in \mathcal{A}^{{\rm LR}}_{\Phi}(B)$, we designate by $\alpha_A : \Lambda  \to \ere  $
the function $\alpha_A(x) := A(x) \cdot x , \forall x \in \Lambda.$ 
Let $\delta > 1$. We denote by $ \mathcal{A}_{\Phi, \delta}(B) $ the set of vector potentials $A \in \mathcal{A}^{{LR}}_\Phi(B) \cap C^2(\Lambda,\ere^3 )$ such that 
 there is a constant $C$ satisfying  
\begin{align}\label{hvlwso-d.1.e1} 
\sum_{i=1}^3\Big |\frac{\partial }{\partial x_i} A(x) \Big |  & \leq C \,\frac{1}{(1 + |x|)^{2}}, \, |\nabla \alpha_{A}(x) | \leq C \,\iota_{2, \delta}(x), \, |\alpha_{A}(x) | \leq C \,  \iota_{1, \delta}(x),  \notag 
\Big |\frac{\partial}{\partial x_i} \frac{\partial}{\partial x_j} \alpha_{A}(x) 
\Big | & \leq C \, \min \Big( \iota_{3, \delta}(x), \frac{\ln(e + |x|)}{(1 + |x|)^2}\Big), 
\end{align}
for all  $x \in \Lambda\,  \,i,j \in \{ 1, 2,3\}.$ }
\end{definition}

It is not difficult to see (see Lemma 3.8 in \cite{bw}) that for any pair  $A^{(1)}, A^{(2)} \in \mathcal{A}_{\Phi}^{{\rm LR}}(B)$ there exists a
 $C^1 \, 0-$ form $\lambda$ in $\overline{\Lambda}$
such that 
$A^{(2)} - A^{(1)}= \nabla \lambda $.  
$\lambda$ is given by the formula 
\beq \label{claro}
\lambda(x):=\int_{C(x_0,x)}(A^{(2)}-A^{(1)}),
\ene
for a fixed point  $x_0 $ in $\Lambda$ and a $C^\infty$-curve  $C(x_0,x)$ in $\Lambda$ with starting point   $x_0$ and ending point  $x$. Moreover the limit 
\beq \label{lambdainfty} 
 \lambda_\infty(x):=\lim_{r\rightarrow \infty} \lambda(rx) 
\ene 
 exists and defines a continuous and homogeneous (of order zero) function in $  \mathbb{R}^3 \setminus \{ 0 \}  $. Furthermore,
\begin{align}
|\lambda_\infty(x)-\lambda(x)|\leq \int_{|x|}^\infty \, b(|x|), \hbox{\rm for some positive function}\,\, b \in L^1(0,\infty).
\label{2.12}
\end{align}
\subsubsection{The Hamiltonians}\label{hamilt}
As in \cite{bw-kg}, \cite{w2}, \cite{w3} we write the Klein-Gordon equation as a  $2$ by $2$ system that is first order in time, in order that it is a Hamiltonian  equation. 

\subsection{Free Hamiltonian}
 The free Klein-Gordon equation is given by
\beq \label{freekg}
\Big (  i \frac{\partial}{\partial t}  \Big )^2 \phi =  
\Big( {\mathbf p}^2+ m^2 \Big ) \phi, 
\ene
where  $\mo := - i \nabla $ is the momentum operator and
 $m > 0$ is the mass of the particle, and the solution  $\phi $ is a complex valued function defined in $ \mathbb{R}\times  \mathbb{R}^3 $. For the free evolution the electromagnetic potentials are zero and there is no obstacle.

We define the operator $
B_0 : = \Big( {\bf p} ^2 + m^2 \Big )^{1/2} $,
with domain the Sobolev space ${\bf H}^1(\mathbb{R}^3)$. 
Denote by $\mathcal{H}_0$ the Hilbert space 
\beq \label{fhilbertA}
\mathcal{H}_0 : = \dom(B_0) \oplus L^2(\mathbb{R}^3),
\ene
with inner product $
\langle \phi, \psi \rangle_{\mathcal{H}_0} : = \langle B_0 \phi_1, B_0 \psi_1  \rangle + 
 \langle  \phi_2, \psi_2 \rangle $,
for $\phi = (\phi_1, \phi_2)$, $\psi = (\psi_1, \psi_2)$.  Note that the inner product of $\mathcal{H}_0$ is the sesquilinear form associated to the classical field energy of the free Klein-Gordon equation. 

The free Hamiltonian is the self-adjoint operator 
\beq \label{ham0}
H_0 : =  \begin{pmatrix}
0  & i \\ -i B_0^2 & 0 
\end{pmatrix},
 \,  \text{with domain} \,  \dom(H_0) : = \dom(B_0^2) \oplus \dom( B_0) = {\bf H}^2(\mathbb{R}^3) \oplus {\bf H}^1(\mathbb{R}^3).
\ene
The free Klein-Gordon equation \eqref{freekg}  is equivalent to the system
$$
i\frac{\partial}{\partial t}\, \psi= H_0 \psi, \qquad \psi \in \mathcal{H}_0,
$$
 with $\psi_1= \phi$, and $\psi_2= \frac{\partial}{\partial t}\, \phi$.
 
We denote by 
$ F_W : \mathcal{H}_0 \mapsto L^2(\mathbb{R}^3)\oplus  L^2(\mathbb{R}^3)$ 
the unitary operator: 
\beq \label{fwm}
F_W : = \begin{pmatrix}
B_0 & 0 \\ 0 & 1
\end{pmatrix}.
\ene
One can easily verify that $ 
F_W H_0 F_W^{-1} = B_0 \beta,$
with $
\beta : = \begin{pmatrix}  0  & i \\ -i & 0  \end{pmatrix} $.
Set $Q$ and $Q^{-1}$ the unitary matrices that diagonalize $\beta$:  
\beq \label{fw3}
Q \beta  Q^{-1} = \begin{pmatrix}
1 & 0 \\ 0 & - 1
\end{pmatrix},
\ene
where 
$  Q : = 2^{-1/2} \begin{pmatrix}
1 & i \\ 1  & -i
\end{pmatrix} $. 
We finally define  the unitary operator $U: \mathcal{H}_0  \mapsto L^2(\mathbb{R}^3)\oplus  L^2(\mathbb{R}^3) $,
$U : = Q F_W $. It follows that
\beq \label{diagu}
\hat{H}_0:=U  H_0U^{-1} =  B_0  \begin{pmatrix}
1 & 0 \\ 0 & - 1
\end{pmatrix}.
\ene
In this representation the free Klein-Gordon equation \eqref{freekg} is equivalent to the system,
\beq\label{diag}
i\frac{\partial}{\partial t}\, \psi= \hat{H}_0 \psi, \qquad \psi \in L^2\left( \mathbb{R}^3\right)\oplus  L^2\left( \mathbb{R}^3\right).
\ene
The appropriate position operator, that gives the position of the quantum particle, is multiplication by the variable $x$ in the  diagonal representation of the Klein-Gordon equation  \eqref{diag}. See \cite{bw-kg}, \cite{w2} and \cite{w3} for the issue of the position operator.

\subsection{Interacting Hamiltonian}
The interacting Klein-Gordon equation for a particle in $\Lambda$ is given by,

\beq
\Big (  i \frac{\partial}{\partial t} -  A_0 \Big )^2 \phi =  
\Big( ({\mathbf p}-  A)^2+ m^2 \Big ) \phi,
\label{intkg}
\ene
where  $\phi : \mathbb{R}\times \Lambda \mapsto \mathbb{C}$ is the wave function. As in the free case we formulate \eqref{intkg} as a $2$ by $2$ system that is first order in time. 

We denote by  $  B(\underline A)^2 : = \Big( {\bf p} - A \Big )^2 + m^2  - A_0^2 $. In Subsection 3.2 in \cite{bw-kg} we prove that  $ B(\underline A)^2$ has a realization as a selfadjoint operator in $L^2(\Lambda)$ and that      $ B(\underline A)^2 \geq \varepsilon$ with $\varepsilon$ as in \eqref{pos}. We designate by

 \beq \label{hilbertA}
\mathcal{H}(\underline A) : = \dom(B(\underline A)) \oplus L^2(\Lambda)
\ene
the Hilbert space with inner product: $
\langle \phi, \psi \rangle_{\mathcal{H}(\underline A)} : = \langle B(\underline A)\phi_1, B(\underline A)\psi_1  \rangle + 
 \langle  \phi_2, \psi_2 \rangle ,  $
for $\phi = (\phi_1, \phi_2)$, $\psi = (\psi_1, \psi_2)$. The inner product of $\mathcal{H}(\underline{A})$ is the sesquilinear form associated to the classical field energy of the interacting Klein-Gordon equation \eqref{intkg}.

The interacting Hamiltonian is the self-adjoint operator (see Subsection 3.2 in \cite{bw-kg}), defined in $\mathcal{H}(\underline A)$,  
\beq \label{ham}
H(\underline{A}) : =  \begin{pmatrix}
0  & i \\ -i B(\underline A)^2 & 2 A_0
\end{pmatrix},
\ene
with domain $  \dom(H(\underline A)) : = \dom(B(\underline A)^2) \oplus \dom( B(\underline A)) $. 

The interacting Klein-Gordon equation \eqref{intkg}  is equivalent to the system,
$$
i\frac{\partial}{\partial t}\, \psi= H(\underline{A}) \,\psi, \qquad \psi \in \mathcal{H}(\underline{A}),
$$
 with $\psi_1= \phi$, and $\psi_2= \frac{\partial}{\partial t}\, \phi$.
 
\section{Wave and Scattering Operators}\label{wavescat}\sss

\subsection{Wave Operators}\label{wave}

The wave operators are defined as follows:
\beq
  W_{\pm}(\underline A):= \hbox{\rm s-}\lim_{t \rightarrow \pm \infty} e^{it
H(\underline A)}\, J\, e^{-it H_0},
\label{2.19}
\ene
provided that the strong limits exist. Here,
 \beq\label{J}
J : = \begin{pmatrix} 
B(\underline A)^{-1} \chi_\Lambda B_0 & 0 \\ 0 & \chi_\Lambda 
\end{pmatrix},
\ene 
is a bounded identification operator from $\mathcal H_0$ into $\mathcal H(\underline A).$

\subsubsection{Existence of the Wave Operators}\label{existence-wave}
In this section we prove existence of wave operators \eqref{2.19} for every magnetic potential $A \in \mathcal{A}^{{\rm LR}}_\Phi(B)$. We provide additionally a change of gauge formula. 

We  first state a Lemma we that use, it is proved in  Lemma 3.26 in \cite{bw-kg}.
\begin{lemma} \label{pento}
We denote by ${\bf v} : \mathbb{R}^3  \mapsto \mathbb{R}^3$,
\beq \label{vel}
{\bf v}(p ) : = \frac{p}{\left(p^2 + m ^2\right)^{1/2}},
\ene
the function that associates to each momentum $p$ the corresponding velocity. 
Take $x_0 \in \mathbb{R}^3 \setminus \{ 0 \} $, $r_0 \in (0, \frac{1}{2}|x_0|)$ and $f \in {\mathcal S}(\mathbb{R}^3)$  be such that $ {\bf v} \big( \operatorname{supp} ( f)\big ) \subset   B(x_0; r_0 )$. For every $l\in \mathbb{N}$ there is a constant $C_l$
such that 
\beq\label{masta}
\Big \|  \chi_{ \big ( B(0; |t|r_0/2 ) + t B(x_0; r_0)\big )^{\displaystyle c }}\, e^{- i t B_0} f(\mo)
\chi_{B \big (0;  |t| r_0/2\big )}   \Big \| \leq C_{l} (1 + |t|)^{-l }, 
\ene
where $\chi_{O}$ is the characteristic function of the set $O$.  
Moreover, let $\tau \in C_0^\infty(\mathbb{R}^3; [0, 1])$ be such that 
$\tau(x) = 1$ for $|x| \leq \frac{1}{2} $ and it vanishes for $ |x| \geq 1 $.  There exists 
$v_0 > 0$ and a constant $C_l$, for every $l \in \mathbb{N}$, such that
\beq\label{nonsta}
\Big \|  \chi_{ B(\nu t; |t|/2 )^{\displaystyle c }}\, e^{- i t B_0} \tau \Big ( \frac{16(\mo - v \nu)}{v }\Big ) 
\chi_{B(0; |t|/8)}   \Big \| \leq C_{l} (1 + |t|)^{- l}, 
\ene
for every $\nu  \in \mathbb{S}^2$ and every $v \geq v_0$.
\end{lemma}

\begin{lemma} \label{TL2}
Let $ A  = \nabla \lambda 
\in \mathcal{A}^{{\rm LR}}_{0}(0)$. Extend $\lambda $ to a $C^1$ function in 
$\mathbb{R}^3$, without changing notation. Set $\lambda_\infty$ as in \eqref{lambdainfty}. Then,
\beq \label{TL21}
s-\lim_{t \to \infty} e^{\pm i\lambda_\infty( \mo/(\mo^2 + m^2)^{1/2} + {\bf x}/t  )} =  
e^{\pm i \lambda_\infty( \mo/(\mo^2 + m^2)^{1/2}   )} = 
 e^{\pm i \lambda_\infty( \mo)},
\ene
where the strong limit is taken in $L^2(\mathbb{R}^3).$ We recall that ${\bf x}$ is the multiplication-by-$x$ operator.
\end{lemma}
\noindent
\emph{Proof:}
We prove the assertion taking the minus sign, the proof with the plus sign is the same.
The second equality is obvious because $\lambda_\infty$ is homogeneous of degree $0$. 
The spectral measure of the operator ${\bf x}/t$ is the projection-valued measure $P_{{\bf x}/t}$ that associates to each Borel set
$\mathcal{B}$   
$P_{{\bf x}/t}(\mathcal{B}) = \chi_{t \mathcal{B}}({\bf x}), \, x \in \mathbb R^3.$ 
As 
\beq \label{PR2}
 \mo/(\mo^2 + m^2)^{1/2} + {\bf x}/t  =  e^{it B_0}{\bf x}/t e^{- it B_0},
\ene
the corresponding spectral measure of this operator is given by  
$P_{ \mo/(\mo^2 + m^2)^{1/2} + {\bf x}/t }(\mathcal{B}) =  e^{it B_0}\,\chi_{t \mathcal{B}} \, e^{- it B_0},$
 for every Borel set $\mathcal{B}$. Let $\phi \in {\mathcal S}(\mathbb{R})^3 $ with $\widehat \phi f = \widehat{\phi}$, with $f$ satisfying the hypotheses of in Lemma \ref{pento}. Then, using \eqref{masta} and the decay of $\phi$, we prove that
\begin{align} \label{PR4}
\| P_{ \mo/(\mo^2 + m^2)^{1/2} + {\bf x}/t }( B(0; r_0/4)) \phi \| 
 = \|   \chi_{t B(0; r_0/4)} e^{- it B_0} 
  f(\mo)( 1 - \chi_{B(0; |t|r_0/2 )} + \chi_{B(0; |t|r_0/2 )} )   \phi   \| \leq C \frac{1}{1 + |t|}.
\end{align}
Let $g_{r_0} : \mathbb{R}^3 \to \mathbb{C}$ be a continuous, bounded by 1, function that equals $e^{- i \lambda_\infty}$ in the complement of $ B(0; r_0/4)$. Eq. \eqref{PR4} implies that 
\beq \label{PR5}
 \lim_{t \to \infty}\Big \| \Big [e^{- i \lambda_\infty( \mo/(\mo^2 + m^2)^{1/2} + {\bf x}/t)} - g_{r_0}\big (\mo/(\mo^2 + m^2)^{1/2} + {\bf x}/t\big ) \Big ] \phi  \Big \| = 0.
\ene
As $g_{r_0}$ is continuous and  $  \mo/(\mo^2 + m^2)^{1/2} + {\bf x}/t  \to \mo/(\mo^2 + m^2)^{1/2}  $ in the strong resolvent sense, Theorem VIII.20 in \cite{rs1} implies that 
\beq \label{PR6}
 s-\lim_{t \to \infty}  g_{r_0}\big (\mo/(\mo^2 + m^2)^{1/2} + {\bf x}/t\big ) =   g_{r_0}\big (\mo/(\mo^2 + m^2)^{1/2} \big ).
\ene
Varying $x_0$, $r_0$ and $g_{r_0}$ and using \eqref{PR5}-\eqref{PR6} we obtain Eq. \eqref{TL21}. 
\bull

\begin{lemma} \label{TL3}
Set 
$ A^{(1)},  A^{(2)} \in  \mathcal{A}^{{\rm LR}}_{\phi}(B)$. Take $\lambda \in \mathbb{C}^1(\mathbb{R}^3)$ such that $A^{(2)}(x) - A^{(1)}(x) = \nabla \lambda (x) $, $x \in \overline{\Lambda}$  [see \eqref{claro}]. Then
\begin{align}\label{LR0}
\mathrm{s-}\lim_{t \to \pm \infty }\Big [ B_0  \kappa'({\bf x})e^{-i  \lambda({\bf x}) } B_0^{-1} -  e^{-i  \lambda_\infty(\pm \mo ) } \Big ]
e^{-it B_0}  = & \,\,0, \,\,\, 
\mathrm{s-}\lim_{t \to \pm \infty }\Big [  \kappa'({\bf x})e^{-i  \lambda({\bf x}) }  -  e^{-i  \lambda_\infty(\pm \mo ) } \Big ]
e^{-it B_0}  =  0
\end{align}
 here the strong limit is taken in $L^2(\mathbb{R}^3)$ and $\kappa' \in C^{\infty}(\mathbb{R}^3)$ is such that $1 - \kappa'$ is compactly supported. Recall the ${\bf x}$ is the multiplication-by-$x$ operator.    
\end{lemma}
\noindent{\it Proof:} We prove assertion  using the plus sign. The proof with the minus sign is the same. We only prove  the first equation in \eqref{LR0}, which is the difficult part (proving the second uses the similar arguments).   Take  $\phi \in {\mathcal S}(\mathbb{R}^3)$ such that $ f(p) \hat \phi(p) = \hat \phi(p) $, for some $f$ satisfying the hypotheses of Lemma \ref{pento}. \\
We have that 
\begin{eqnarray}  \label{LR01}
\notag \lim_{t \to  \infty } \Big \|  B_0 \Big [   \kappa'({\bf x})e^{-i  \lambda({\bf x}) } B_0^{-1} -  B_0^{-1}  e^{-i  \lambda_\infty( \mo ) } \Big ]
e^{-it B_0} \phi \Big \| & \leq & C  \lim_{t \to  \infty } \Big \|  \Big [   \kappa'({\bf x})e^{-i  \lambda({\bf x}) } B_0^{-1} -  B_0^{-1}  e^{-i  \lambda_\infty( \mo ) } \Big ]
e^{-it B_0} \phi \Big \|_{{\bf H}^1(\mathbb{R}^3) }\\
 \leq  C \lim_{t \to  \infty } \Big \|   \big [\mo, \,  \kappa'({\bf x}) e^{-i  \lambda({\bf x}) } \big ] B_0^{-1}  
e^{-it B_0} \phi \Big \|
   + &C& \lim_{t \to  \infty } \Big \|  \Big [    \kappa'({\bf x}) e^{-i  \lambda({\bf x}) }   -   e^{-i  \lambda_\infty( \mo ) } \Big ]
e^{-it B_0} \mo B_0^{-1} \phi \Big \| \\ \notag
    + C \lim_{t \to  \infty } \Big \|  \Big [    \kappa'({\bf x}) e^{-i  \lambda({\bf x}) }   -   e^{-i  \lambda_\infty( \mo ) } \Big ]
e^{-it B_0}  B_0^{-1} \phi \Big \|. 
\end{eqnarray}

As  $  \nabla \lambda (x) =  A^{(2)}(x) - A^{(1)}(x)  $ decays as $\frac{1}{|x|}$ as $|x|$ tends to infinity, the commutator 
$  \big [\mo, \,  \kappa'({\bf x}) e^{-i  \lambda({\bf x}) } \big ] $ decays as 
$\frac{1}{|x|}$ as $|x|$ tends to infinity, which together with Lemma \ref{pento} (or just the Rellich-Kondrakov lemma) imply  that
\begin{align} \label{LR02}
\mathrm{s-}\,\lim_{t \to  \infty }    \big [\mo, \,  \kappa'({\bf x}) e^{-i  \lambda({\bf x}) } \big ] B_0^{-1}  
e^{-it B_0} \phi = 0. 
\end{align}   
Then, we obtain, using \eqref{LR01}-\eqref{LR02}, that \eqref{LR0} is valid whenever 
\begin{align}\label{LR03}
\mathrm{s-}\, \lim_{t \to  \infty }\Big [   \kappa'({\bf x})e^{-i  \lambda({\bf x}) }  -  e^{-i  \lambda_\infty( \mo ) } \Big ]
e^{-it B_0} \phi = \mathrm{s-}\,\lim_{t \to  \infty } e^{it B_0} \Big [   e^{-i  \lambda({\bf x}) }  -  e^{-i  \lambda_\infty( \mo ) } \Big ]
e^{-it B_0} \phi = 0, 
\end{align}
in $L^2(\mathbb{R}^3)$, for every $\phi$ as above.
We use Lemma \ref{pento} (or just the Rellich-Kondrakov lemma) to prove that
\begin{align}\label{LR04}
\mathrm{s-}\,\lim_{t \to  \infty } (1 -   \kappa'({\bf x}))e^{-i  \lambda({\bf x}) } 
e^{-it B_0} \phi = 0,
\end{align}  
which implies the first equality in \eqref{LR03}.  
Additionally, Lemma \ref{pento}  and the decay of $\phi$ imply that (here we use the notation of the referred lemma) 
\begin{align} \label{LR05}
\mathrm{s-}\,\lim_{t \to  \infty } \Big [   e^{-i  \lambda({\bf x}) }  -  e^{-i  \lambda_\infty( {\bf x} ) } \Big ]
e^{-it B_0} \phi = \mathrm{s-}\,\lim_{t \to  \infty }  \chi_{ \big ( B(0; |t|r_0/2 ) + t B(x_0; r_0)\big )}  \Big [   e^{-i  \lambda({\bf x}) }  -  e^{-i  \lambda_\infty( {\bf x} ) } \Big ]
e^{-it B_0} \phi .
\end{align}
Eq. \eqref{2.12} implies that there is a positive decreasing function $h : [0, \infty) \mapsto [0,  \infty) $ with $\lim_{r \to \infty } h(r) =0$, such that 
\begin{align}\label{LR06}
 \Big \| \chi_{ \big ( B(0; |t|r_0/2 ) + t B(x_0; r_0)\big )}  \Big [   e^{-i  \lambda({\bf x}) }  -  e^{-i  \lambda_\infty( {\bf x} ) } \Big ] \Big \|
 \leq h(t),
\end{align}
from which, together with \eqref{LR05}, we get 
\begin{align} \label{LR07}
\mathrm{s-}\,\lim_{t \to  \infty }e^{it B_0} \Big [   e^{-i  \lambda({\bf x}) }  -  e^{-i  \lambda_\infty( {\bf x} ) } \Big ]
e^{-it B_0} \phi = 0.
\end{align}  
Using \eqref{PR2} and the fact that $\lambda_\infty $ is homogeneous of degree $0$ we prove
\begin{equation}\label{LR08}
\mathrm{s-}\,\lim_{t \to  \infty }e^{it B_0}  e^{-i  \lambda_\infty( {\bf x} ) } 
e^{-it B_0} \phi = \mathrm{s-}\,\lim_{t \to  \infty }e^{it B_0}  e^{-i  \lambda_\infty( {\bf x}/t ) } 
e^{-it B_0} \phi =\mathrm{s-}\, \lim_{t \to  \infty } 
 e^{-i  \lambda_\infty\big ( \mo/(\mo^2 + m^2)^{1/2}  +  {\bf x}/t \big) }  \phi = e^{-i  \lambda_\infty\big ( \mo \big) }  \phi,
\end{equation} 
where we used Lemma \ref{TL2}. Eqs. \eqref{LR04}, \eqref{LR07} and \eqref{LR08} imply \eqref{LR03}, which in turn implies the desired result.

\begin{theorem}[Existence of Wave Operators and Change of Gauge Formula] \label{exchan}
For every $ \underline A = (A_0, A)$  with  $ A \in A^{{\rm LR}}_{\Phi}(B)$  the limits \eqref{2.19} exist and are isometric. For every 
$ \underline A^{(i)} = (A_0, A^{(i)})$, $i \in \{1, 2 \}$, with 
$ A^{(1)},  A^{(2)} \in  \mathcal{A}^{{\rm LR}}_{\phi}(B)$
\beq \label{changeW}
W_{\pm}(\underline A^{(2)}) = e^{ i \lambda({\bf x}) }W_{\pm}(\underline A^{(1)})  U^{-1 }   \begin{pmatrix}
  e^{- i \lambda_\infty( \pm\mo )}  & 0 \\ 
 0 &  
  e^{- i \lambda_{\infty}(\mp \mo)}
\end{pmatrix}  U,
\ene
 where $A^{(2)} - A^{(1)} = \nabla \lambda $ [see \eqref{claro}]. Notice that in the expression above  $ e^{ i \lambda({\bf x}) }$ is an operator from  $ \mathcal{H}(\underline{A}^{(1)})  $  to $ \mathcal{H}(\underline{A}^{(2)})   $.   
\end{theorem}
\noindent \emph{Proof:}
We suppose that $A^{(1)} \in \mathcal{A}^{{\bf SR}}_\Phi(B)$ (recall that by Remark \ref{coupot} the set  $ \mathcal{A}_{\Phi}^{{\rm SR}}(B) $ is not empty) . Theorem 3.4 and the proof of Lemma 3.2  in 
\cite{bw-kg} assure that  the wave operators $  W_{\pm}(\underline A^{(1)}) $ exist,  are isometric and that
$$
  W_{\pm}({\underline A}^{(1)}):= \hbox{\rm s-}\lim_{t \rightarrow \pm \infty} e^{it
H({\underline A}^{(1)})}\, \kappa\, e^{-it H_0},
$$
for any $\kappa$ as in \eqref{2.4} and the text above it. By proving the change of gauge formula we prove the existence of $  W_{\pm}(\underline A^{(2)}) $. See Lemma 3.2 of \cite{bw-kg} whose proof applies also in this case. Clearly, the existence and the isometry of the wave operators, and the change of gauge formula in the case of general  $A^{(1)}$ and $A^{(2)}$ follows from the same result in the case when one of the potentials is in $\mathcal{A}^{{\bf SR}}_\Phi(B)$ .  We prove the assertion for $W_{+}$. The proof for $ W_{-}$ is analogous. A simple computation gives 
\beq \label{PRW1} 
H(\underline A^{(2)} ) =  e^{i \lambda ({\bf x})} H(\underline A^{(1)} )e^{-i \lambda ({\bf x})}, 
\ene 
which implies that
\beq \label{PRW2}
W_+(\underline A^{(2)}) = e^{i \lambda ({\bf x})} \mathrm{s-} \lim_{t \to \infty} e^{it H(\underline A^{(1)})} \kappa e^{-it H_0} 
e^{it H_0} \kappa' e^{- i \lambda({\bf x})} e^{-it H_0},   
\ene
whenever the limit exists. Here $\kappa' $ satisfies the properties of $\kappa$ (i.e. it satisfies  \eqref{2.4} and the text above it). Additionally, we suppose that
$\kappa' \kappa = \kappa$.
Using \eqref{diagu} we obtain that  
\begin{align} \label{PRW5}
e^{it H_0} \kappa' e^{- i \lambda({\bf x})} e^{-it H_0} = & U^{-1 }  \begin{pmatrix}
e^{it B_0} & 0 \\ 0 & e^{-it B_0}
\end{pmatrix} Q  \begin{pmatrix}
 B_0 \kappa' e^{- i \lambda({\bf x})}B_0^{-1} & 0 \\ 0 &  \kappa' e^{- i \lambda({\bf x})}
\end{pmatrix}     Q^{-1 }  \begin{pmatrix}
e^{-it B_0} & 0 \\ 0 & e^{it B_0}
\end{pmatrix} U  \\ \notag  = &  \frac{1}{2}U^{-1 }  \begin{pmatrix}
e^{it B_0} & 0 \\ 0 & e^{-it B_0}
\end{pmatrix}   \begin{pmatrix}
 B_0 \kappa' e^{- i \lambda({\bf x})}B_0^{-1} +  \kappa' e^{- i \lambda({\bf x})} & 
 B_0 \kappa' e^{- i \lambda({\bf x})}B_0^{-1} -  \kappa' e^{- i \lambda({\bf x})} \\ 
 B_0 \kappa' e^{- i \lambda({\bf x})}B_0^{-1} -  \kappa' e^{- i \lambda({\bf x})} &  
 B_0 \kappa' e^{- i \lambda({\bf x})}B_0^{-1} +  \kappa' e^{- i \lambda({\bf x})}
\end{pmatrix} 
\\ & \hspace{10cm} \notag \cdot     
 \begin{pmatrix}
e^{-it B_0} & 0 \\ 0 & e^{it B_0}
\end{pmatrix} U, 
\end{align}
which together with Lemma \ref{TL3} imply that 
\begin{align}\label{PRW50}
s-\lim_{t \to \infty}  e^{it H_0} \kappa' e^{- i \lambda({\bf x})} e^{-it H_0}  =   U^{-1 }   \begin{pmatrix}
  e^{- i \lambda_\infty(\mo )}  & 0 \\ 
 0 &  
  e^{- i \lambda_{\infty}(-\mo)}
\end{pmatrix}  U, 
\end{align}
from which the desired result follows.
\bull

 \subsection{Scattering Operator} \label{existence-scattering}
The scattering operator is defined, for every $\underline A = (A_0, A)$ (with 
$A \in \mathcal{A}_{\Phi}^{{\rm LR}}(B)$) by 
\begin{equation}
S(\underline A) = W_+^*(\underline{A})W_-(\underline{A}).
\end{equation}
The following theorem gives the change of gauge formula for the scattering operator.   
\begin{theorem}\label{changescat}
For every 
$ \underline A^{(i)} = (A_0, A^{(i)})$, $i \in \{1, 2\}$, with 
$ A^{(1)},  A^{(2)} \in  \mathcal{A}^{{\rm LR}}_{\Phi}(B)$,
\beq \label{change}
S(\underline A^{(2)}) =  
  U^{-1 }   \begin{pmatrix}
  e^{ i \lambda_\infty(\mo )}  & 0 \\ 
 0 &  
  e^{ i \lambda_{\infty}(-\mo)}
\end{pmatrix}  U S(\underline A^{(1)})  U^{-1 }   \begin{pmatrix}
  e^{- i \lambda_\infty( -\mo )}  & 0 \\ 
 0 &  
  e^{- i \lambda_{\infty}(\mo)}
\end{pmatrix}  U, 
\ene
 where $A^{(2)} - A^{(1)} = \nabla \lambda $ [see \eqref{claro}]. 
\end{theorem}
\noindent {\it Proof:} The result is a direct consequence of Theorem \ref{exchan}. Notice that the dual of the operator $ e^{ i \lambda({\bf x}) } : \mathcal{H}(\underline{A}^{(1)})  \mapsto  \mathcal{H}(\underline{A}^{(2)})   $ is  $ e^{ -i \lambda({\bf x}) } : \mathcal{H}(\underline{A}^{(2)})  \mapsto  \mathcal{H}(\underline{A}^{(1)})  $. 
\bull 

In \cite{bw} we considered change of gauge formulae where the fluxes can differ in multiples of  $2 \pi$, in the case of the Schr\"odinger equation. Similar results are true for the Klein-Gordon equation.  
 
\section{High Momenta  Limit of the Scattering Operator}\label{high-long} \sss
In this section we prove one of our main results: We give a high-momenta expression for the scattering operator, with error bounds. This formula is the content of Theorem \ref{TPG}, which is a generalization of Theorem 2.8 in \cite{bw-kg} to long-range magnetic potentials. Our formula is used to reconstruct important information from the potentials and the magnetic field. \\

For every $ \nu \in \mathbb{S}^1$ we denote by (see Eq. \eqref{2.4})  
\beq 
\Lambda_{ \nu} : = \Big \{ x \in \Lambda \, : \, x + \tau \nu \in \Lambda, \:  \forall \tau \in \mathbb{R} \Big \}, 
\, \mathrm{and}\,
\Lambda_{\kappa, \nu} : = 
\Big \{x \in \Lambda_{\nu} \, : \, \kappa(x + \tau \nu) = 1, \:  \forall \tau \in \mathbb{R}  \Big   \}. 
\ene   


\begin{theorem} \label{TPG}
Set $\nu \in  \mathbb{S}^2 $ and $l \in \mathbb{N}$,  $l \geq \zeta/2, l \geq 2$.
Suppose that $ \phi, \psi \in  {\bf H}_{\langle x \rangle^{4 l}}^2(\mathbb{R}^3)^2$
are supported in $\Lambda_{\kappa, \nu} $. Let $ \underline A = (A_0, A)$,  with  $ A \in A^{{\rm LR}}_{\Phi}(B)$. Suppose that for some  $\zeta_0 >1$  there is $ A^{(1)} \in A^{{\rm SR}}_{\Phi}(B)$ satisfying \eqref{noq}  with $\zeta=\zeta_0$. 
Then  
\begin{align} \label{TP1I}
\langle U  S (\underline{A})  U^{-1} e^{i {\bf x} \cdot v\nu }\phi \; ,  e^{i {\bf x} \cdot v \nu } 
\psi  \rangle_{L^2(\mathbb{R}^3)^2} =  &   \Big \langle  \begin{pmatrix}
 e^{i \int_{-\infty}^{\infty} dr  (   A \cdot \nu - A_0) ({\bf x} + r \nu ) }  & 0 
\\ 0 & e^{- i \int_{-\infty}^{\infty} dr   ( A \cdot \nu + A_0) ({\bf x} + r \nu ) }
\end{pmatrix}  \phi,\; \psi \Big \rangle  ,
 \\ & +  \|    \phi   \|_{ {\bf H}_{\langle x \rangle^{4 l}}^2 (\mathbb{R}^3)}  
 \|    \psi   \|_{{\bf H}_{\langle x \rangle^{4 l}}^2 (\mathbb{R}^3)}
 \begin{cases} O\Big ( v^{1 - \zeta_0}  + \frac{1}{v} \Big ), & \text{if} 
 \: \zeta_0 \ne 2, \\ \\
 O\Big ( \frac{ \ln(v)}{v}  \Big ), & \text{if} 
\: \zeta_0 = 2. 
 \end{cases} \notag
\end{align}

\end{theorem}

\noindent Recall that by Remark \ref{coupot} the set $A^{{\rm SR}}_{\Phi}(B)$ is not empty, i.e. there is always a potential (the Coulomb potential) in  $A^{{\rm SR}}_{\Phi}(B)$ that satisfies \eqref{noq} for some $\zeta >1$ that depends on the decay rate, $\mu$, of the magnetic field. See \eqref{2.2}.

\noindent{\it Proof of the Theorem: } 
We identify $A^{(2)} \equiv  A \in \mathcal{A}^{{\rm LR}}_\Phi(B) $. Set $\lambda$ such that $ A^{(2)} - A^{(1)} = \nabla \lambda  $ and take $\underline A^{(j)} = (A_0, A^{(j)}) $, $j  \in \{1, 2  \}$. 
By the fact that $ \lambda_\infty $ is homogeneous of degree $0$ and  Lemma 3.8 in \cite{bw} 
\begin{align}\label{bw38}
|\lambda_\infty(x + y) -  \lambda_\infty(x)| \leq C |y|
\end{align}
for every $x \in \mathbb{R}^3\setminus \{ 0\}  $ and every $y$ with $|y | \leq \frac{|x|}{2} $. Take $\tau \in C_0^\infty(\mathbb{R}^3; [0, 1])$ be such that 
$\tau(x) = 1$ for $|x| \leq \frac{1}{2} $ and it vanishes for $ |x| \geq 1 $. We have that, for every  $\varphi \in {\bf H}^1(\mathbb{R}^3)$, 
\begin{align}\label{tet1}
\Big \| \big [ e^{- i {\bf x}\cdot  v \nu } e^{i \lambda_\infty (\mo)} e^{i {\bf x} \cdot v \nu  }  -  e^{i \lambda_\infty (\nu)} \big ]  \varphi \Big  \|
\leq & \Big \| \big [ e^{i \lambda_\infty (\mo/v + \nu )}   -  e^{i \lambda_\infty (\nu)} \big ] \tau(2 \mo /v) \varphi \Big  \| +
2 \Big \| (1 - \tau(2 \mo /v)) \frac{1}{\langle \mo \rangle}\Big \|   \| \varphi   \|_{{\bf H}^1(\mathbb{R}^3)}
\\ \notag  \leq & C \frac{1}{v}  \| \varphi   \|_{{\bf H}^1(\mathbb{R}^3)},
\end{align}
where we use \eqref{bw38}. 
By Theorem \ref{changescat}, \eqref{tet1} and  arguments similar to the ones used in the proof of \eqref{tet1}, we obtain
\begin{align} \label{changed}
\langle e^{-i {\bf x} \cdot v \nu  } & U S(\underline A^{(2)}) U^{-1} e^{i{\bf x} \cdot v \nu} \phi , \psi  \;  \rangle \\ \notag &  =  
 \Big \langle e^{-i {\bf x} \cdot v \nu  }   \begin{pmatrix}
  e^{ i \lambda_\infty(\mo )}  & 0 \\ 
 0 &  
  e^{ i \lambda_{\infty}(-\mo)}
\end{pmatrix}  U S(\underline A^{(1)})  U^{-1 }   \begin{pmatrix}
  e^{- i \lambda_\infty( -\mo )}  & 0 \\ 
 0 &  
  e^{- i \lambda_{\infty}(\mo)}
\end{pmatrix}  e^{i{\bf x} \cdot v \nu} \phi , \psi  \; \Big \rangle
\\ \notag &  =  
 \Big \langle    \begin{pmatrix}
  e^{ i \lambda_\infty(\nu )}  & 0 \\ 
 0 &  
  e^{ i \lambda_{\infty}(-\nu)}
\end{pmatrix}  e^{-i {\bf x} \cdot v \nu  } U S(\underline A^{(1)}) U^{-1}  e^{i{\bf x} \cdot v \nu}  \begin{pmatrix}
  e^{- i \lambda_\infty( -\nu )}  & 0 \\ 
 0 &  
  e^{- i \lambda_{\infty}(\nu)}
\end{pmatrix}  \phi , \psi  \; \Big \rangle
\\ & + \| \phi   \|_{{\bf H}^1(\mathbb{R}^3)^2}\| \psi   \|_{{\bf H}^1(\mathbb{R}^3)^2} O\Big ( \frac{1}{v}\Big ) \notag.
\end{align} 
Using Theorem 2.8 in \cite{bw-kg} we obtain
\begin{align} \label{TP1tar} 
\Big \langle     \begin{pmatrix}
  e^{ i \lambda_\infty(\nu )}  & 0 \\ 
 0 &  
  e^{ i \lambda_{\infty}(-\nu)}
\end{pmatrix} &  e^{-i {\bf x} \cdot v \nu } U S (\underline{A}^{(1)})  U^{-1} e^{i {\bf x} \cdot v\nu }   \begin{pmatrix}
  e^{- i \lambda_\infty( -\nu )}  & 0 \\ 
 0 &  
  e^{- i \lambda_{\infty}(\nu)}
\end{pmatrix}   \phi \;  , 
\psi  \Big \rangle_{L^2(\mathbb{R}^3)^2} 
\\ \notag   &  \hspace{-3cm} = \langle      \begin{pmatrix}
  e^{ i \lambda_\infty(\nu )}  & 0 \\ 
 0 &  
  e^{ i \lambda_{\infty}(-\nu)}
\end{pmatrix}   \begin{pmatrix}
 e^{i \int_{-\infty}^{\infty} dr  (   A^{(1)} \cdot \nu - A_0) ({\bf x} + r \nu ) }  & 0 
\\ 0 & e^{- i \int_{-\infty}^{\infty} dr   ( A^{(1)} \cdot \nu + A_0) ({\bf x} + r \nu ) }
\end{pmatrix}  \begin{pmatrix}
  e^{- i \lambda_\infty( -\nu )}  & 0 \\ 
 0 &  
  e^{- i \lambda_{\infty}(\nu)}
\end{pmatrix}  
\phi \; , 
  \psi  \rangle_{L^{2}(\mathbb{R}^3)^2} 
 \\ &  \hspace{4cm} +  \|    \phi   \|_{\bf H_{\langle x \rangle^{4 l}}^2 (\mathbb{R}^3)^2}  
 \|    \psi   \|_{\bf H_{\langle x \rangle^{4 l}}^2 (\mathbb{R}^3)^2}
 \begin{cases} O\Big ( v^{1 - \zeta_0}  + \frac{1}{v} \Big ), & \text{if} 
 \: \zeta_0 \ne 2, \\ \\
 O\Big ( \frac{ \ln(v)}{v}  \Big ), & \text{if} 
\: \zeta_0 = 2. \notag
 \end{cases} 
\end{align}
We get the desired result using Eqs. \eqref{changed}-\eqref{TP1tar} and
\begin{align}\label{tet2}
 \begin{pmatrix}
  e^{ i \lambda_\infty(\nu )}  & 0 \\ 
 0 &  
  e^{ i \lambda_{\infty}(-\nu)}
\end{pmatrix} &   \begin{pmatrix}
 e^{i \int_{-\infty}^{\infty} dr  (   A^{(1)} \cdot \nu - A_0) ({\bf x} + r \nu ) }  & 0 
\\ 0 & e^{- i \int_{-\infty}^{\infty} dr   ( A^{(1)} \cdot \nu + A_0) ({\bf x} + r \nu ) }
\end{pmatrix}  \begin{pmatrix}
  e^{- i \lambda_\infty( -\nu )}  & 0 \\ 
 0 &  
  e^{- i \lambda_{\infty}(\nu)}
\end{pmatrix} \\ \notag   &   \hspace{5cm}= 
   \begin{pmatrix}
 e^{i \int_{-\infty}^{\infty} dr  (   A^{(2)} \cdot \nu - A_0) ({\bf x} + r \nu ) }  & 0 
\\ 0 & e^{- i \int_{-\infty}^{\infty} dr   ( A^{(2)} \cdot \nu + A_0) ({\bf x} + r \nu ) }
\end{pmatrix} .
\end{align}

\section{Reconstruction Methods: The magnetic Field, the Electric Potential and Magnetic Fluxes Modulo $2 \pi$}\sss\label{smfepf}
In this section we use the high momenta limit of the scattering operator in Theorem \ref{TPG} to reconstruct the electric potential, the magnetic field and certain fluxes of the magnetic potential around handles of the obstacle. These results extend the results in \cite{bw-kg} (Theorems 2.10 and 2.12), where they are proved only for short-range magnetic potentials. The main new ingredient that allows us to extend the results to the long-range case is Theorem \ref{TPG}. Once it is established, the proofs follows the same lines for both cases. Since they are already presented in \cite{bw-kg}, for the short-range case, we only state the results and refer to \cite{bw-kg} for the proofs. 

\begin{definition}{\rm
We denote by $\Lambda_{{\rm Rec}}$ the set of points $ x \in \Lambda $ such that, for some two-dimensional plane $P_x$, $x + P_x \subset \left(\kappa^{-1}(\{1\})\right)^o $, for some function $\kappa \in C_\infty(\mathbb{R}^3) $ satisfying \eqref{2.4} and the text above it.  }
 
\end{definition}

\begin{theorem} \label{inverse-fields}
The high-momenta limit \eqref{TP1I} of the scattering operator uniquely determines $B(y)$ and $A_0(y)$ for every $y \in \Lambda_{Rec}$. 
\end{theorem}
\noindent{\it  Proof:}
The proof follows the lines of the proof of Theorem 2.10 in \cite{bw-kg}, using Theorem \ref{TPG}. Note that the proof also gives a method for the unique reconstruction of $B(y)$ and $A_0(y)$ for every $y \in \Lambda_{Rec}$. \bull

We denote by
\begin{equation} \label{line}
L(x,\hv):= x+\ere \hv,
\end{equation}
for every $x \in \ere^3$ and any unit vector $\hv \in \ese^2$.  Suppose that
$
L(x,\hv) \cup L(y,\hw) \subset \Lambda, x \cdot y \geq 0,
$
and $ \rho >0$ is such that
$$
\hbox{\rm convex}\, \left( (x+ (-\infty, -\rho] \hv ) \cup  (y+ (-\infty, -\rho] \hw )\right)
\cup \,\hbox{\rm convex}\, \left( (x+ [\rho, \infty) \hv ) \cup  (y+ [\rho, \infty, ) \hw )\right)
\subset \mathbb{R}^3 \setminus {B(0; r)},
$$
where $ K \subset B(0; r)$ (convex$(\cdot)$ denotes the convex hull).
We denote by $\gamma(x,y,\hv,\hw)$ the curve
with sides $ x+[-\rho, \rho]\hv$, oriented in the direction of $\hv$,
 $ y+[-\rho, \rho]\hw$, oriented in the direction of $-\hw$, and the straight lines that join the
 points $x+\rho \hv$ with $y+ \rho\hw$ and $y-\rho \hw$ and $x-\rho \hv$.
\bull

\begin{theorem} \label{th-7.1}
Suppose that $B = 0$ and that $A_0 = 0$. Then, for any flux, $\Phi$,
and all $ A\in \mathcal{A}^{{LR}}_\Phi(0)$, the high-momenta limit of  $S(\underline A$)  in  \eqref{TP1I}, known  for $\hv$ and $\hw$, 
determines the fluxes
\beq
 \int_{\gamma (x,y,\hv,\hw)}A
\label{7.3}
\ene
modulo $2\pi$, for all curves $\gamma (x,y,\hv,\hw)$.
\end{theorem}
\noindent{\it  Proof:}
The proof follows the lines of the proof of Theorem 2.12 in \cite{bw-kg}, using Theorem \ref{TPG}. 
\bull

\section{Reconstruction Methods: Long-Range Magnetic Potentials}\label{rmLR}\sss

In this section we derive information of the long-range part of the magnetic potential from the high momenta limit of the scattering operator. We first introduce some definitions.\\
Take $R >0$ such that
$ K \subset  B(0; R)$. Suppose that $  L(x,\hv) \subset \Lambda$, and
$L(x,\hv) \cap B(0; R) \neq \emptyset$. We denote by  $c(x,\hv) \equiv c_{R, \mathfrak{s}}(x,\hv)$ the curve consisting of the segment
$L(x,\hv) \cap \overline{B(0; R)}$ and  a $C^\infty$ simple differentiable curve $\mathfrak{s}$ on $ \partial\overline{ B(0; R)}$ that connects the points
$L(x,\hv) \cap \partial  \overline{ B(0; R)}$. We orient $c(x,\hv)$ in such a way that the segment of straight
line has the orientation of $\hv$. See Figure 2. We stress that $ c_{R, \mathfrak{s}}(x, \hv) $ depends on the curve we choose joining the points in  $L(x,\hv) \cap \overline{B(0; R)}$. However, the quantities that we associate to it below do not depend on this election. 
 
\begin{definition} \label{long-defi}{\rm
For every magnetic potential $A \in \mathcal{A}^{{\rm LR}}_{\Phi}(B) $ and every 
$\hv  \in \mathbb{S}^2$ we define the quantity $\Phi_L(A, \hv)$, which we name the long-range flux of $A$ in the direction $\hv$, as follows:   Suppose that $ K  \subset B(0; R)$.  Take $x \in \mathbb{R}^3 $ with 
$|x|  \geq  R$ and define the curve $c_{|x|+1, \mathfrak{s}}(x,\hv)$ as above. Then,  denoting $  \mathfrak{s}(s):= s ( \mathfrak{s}-x)+x$,   }

\begin{align}\label{defflu}
\Phi_L(A, \hv) : = 
- \lim_{s \to \infty }
  \int_{ \mathfrak{s}(s)}  A.
\end{align}
\end{definition}
We compute the limit in the right hand side of \eqref{defflu}. Let us  take $ A^{({\rm SR})} \in \mathcal{A}_{\Phi}^{{\rm SR}}(B) $ and $\lambda$ such that $A - A^{({\rm SR})} = \nabla \lambda $. 
. 
The fact that $ A^{({\rm SR})} $ is short-range implies that 
\begin{align} \label{srreg}
\lim_{s \to \infty } \int_{ \mathfrak{s}(s)} A^{({\rm SR})} = 0,
\end{align}
and, therefore,
$$
\lim_{s \to \infty} \int_{s(c_{(|x|+ 1), \mathfrak{s}}( x, \hv)-x)+x} A - A^{({\rm SR})} = 0 =  \int_{L( x, \hv)} \left( A - A^{({\rm SR})}\right) +  
\lim_{s \to \infty}\int_{ \mathfrak{s}(s)} \left( A -  A^{({\rm SR})} \right)
 \notag 
=  \lambda_\infty(\hv) - \lambda_{\infty}(- \hv) + \lim_{s \to \infty }
\int_{ \mathfrak{s}(s)}  A . 
$$
We conclude that 
\begin{align}\label{defflu2}
\Phi_L(A, \hv)  =  \lambda_\infty(\hv) - \lambda_{\infty}(- \hv).
\end{align}
From the definition in Eq. \eqref{defflu} it follows  that $\Phi_L(A, \hv)$ is an intrinsic property of $A$,  although it can also be expressed as $ \lambda_\infty(\hv) - \lambda_{\infty}(- \hv) $,  as \eqref{defflu2} shows.
  It is also clear from \eqref{defflu2} that $\Phi_L(A, \hv)$
 does not depend on the particular curve $ \mathfrak{s} $ that  is used to define it.     

\begin{prop} \label{rem1} 
 Suppose that $ B \in C^{2}(\overline{\Lambda})$ 
     and that $ |  B(x)| \leq C \frac{1}{(1 +|x|)^\mu} $, $ | \frac{\partial}{\partial x_i} B(x)| \leq C \frac{1}{(1 +|x|)^{\mu + 1}} $, $ | \frac{\partial}{\partial x_j}\frac{\partial}{\partial x_i} B(x)| \leq C \frac{1}{(1 + |x|)^{\mu + 2}} $, for every 
$i, j \in \{ 1, 2,3 \}$ and every $x \in \Lambda$.  Let $  \delta > 1$ and  $A \in \mathcal{A}_{\Phi,  \delta }(B)$. Then, 
\begin{align} \label{ainfty}
 A_\infty(\hv):=\lim_{\tau \to \infty}A(\tau \hv)\tau
\end{align} 
   exists, it 
is continuous as a function of $\hv \in \mathbb{S}^2$ and $\forall \hv \in \mathbb{S}^2\: : \: A_\infty (\hv)\cdot \hv = 0$. We extend  \eqref{ainfty} to $\mathbb{R}^3 \setminus \{ 0\}$ taking 
\begin{align}
A_\infty(x) : = \frac{1}{|x|} A_\infty \Big(\frac{x}{|x|}\Big ).
\end{align}
\end{prop}
\noindent {\it Proof:}
The proof follows from Corollary 3.13 in  \cite{bw-2dim}. Although in \cite{bw-2dim} only $2$ dimensions are considered, the proof also applies in our case.  

\begin{prop} \label{rem1.} 
 Suppose that $ B \in C^{2}(\overline{\Lambda})$ 
     and that $ |  B(x)| \leq C \frac{1}{(1 +|x|)^\mu} $, $ | \frac{\partial}{\partial x_i} B(x)| \leq C \frac{1}{(1 +|x|)^{\mu + 1}} $, $ | \frac{\partial}{\partial x_j}\frac{\partial}{\partial x_i} B(x)| \leq C \frac{1}{(1 + |x|)^{\mu + 2}} $, for every 
$i, j \in \{ 1, 2,3 \}$ and every $x \in \Lambda$.  Let $  \delta > 1$ and  $A \in \mathcal{A}_{\Phi,  \delta}(B)$, then for every $\hv \in \mathbb{S}^2$
\begin{align}\label{rem11}
\Phi_L(A, \hv) = & - \int_{ 0 }^{\pi} A_\infty \Big ( \cos(\theta) \hv  + \sin(\theta) \hv^\perp  \Big )\cdot \Big ( - \sin(\theta) \hv  + \cos(\theta) \hv^\perp \Big ) d\theta, \\  \label{rem12}
\frac{\partial}{\partial \vartheta} \Phi_L\Big  (A,  \cos(\vartheta) \hv  + \sin(\vartheta) \hv^\perp  \Big )\Big |_{\vartheta = 0}  = & 
\Big [ A_\infty (\hv) + A_\infty (- \hv) \Big ]\cdot \hv^{\perp}, 
\end{align}
where $\hv^\perp$ is any unit vector,  orthogonal to $\hv$. 
\end{prop}
\noindent{\it Proof:}
Notice that, conveniently selecting $\mathfrak{s}$,   
\begin{align}
\lim_{s \to \infty }\int_{  \mathfrak{s}(s)} A = 
\lim_{s \to \infty } \int_{ 0 }^{\pi} A \Big ( s \cos(\theta) \hv  + s \sin(\theta) \hv^\perp  \Big )\cdot  \Big (  -s \sin(\theta) \hv  + s\cos(\theta) \hv^\perp \Big ) d\theta  
\end{align}
and that, by \eqref{2.10b}, $ A \Big ( s \cos(\theta) \hv  + s \sin(\theta) \hv^\perp  \Big )\cdot  \Big (  s \cos(\theta) \hv  + s \sin(\theta) \hv^\perp \Big ) $ is uniformly bounded (with respect to $s$). Eq. \eqref{rem11} follows from the Lebesgue convergence theorem, \eqref{defflu} and \eqref{ainfty}. Deriving \eqref{rem11} we obtain Eq. \eqref{rem12}.

\begin{lemma}\label{lem}
 Suppose that $ B \in C^{2}(\overline{\Lambda})$ 
     and that $ |  B(x)| \leq C \frac{1}{(1 +|x|)^\mu} $, $ | \frac{\partial}{\partial x_i} B(x)| \leq C \frac{1}{(1 +|x|)^{\mu + 1}} $, $ | \frac{\partial}{\partial x_j}\frac{\partial}{\partial x_i} B(x)| \leq C \frac{1}{(1 + |x|)^{\mu + 2}} $, for every 
$i, j \in \{ 1, 2,3 \}$ and every $x \in \Lambda$.  Let $  \delta > 1$ and  $A \in \mathcal{A}_{\Phi,  \delta}(B)  $, then for $\hv \in \mathbb{S}^2$ the function 
$$
\vartheta \mapsto \int_{L(x,  \cos(\vartheta) \hv  + \sin(\vartheta) \hv^\perp    )} A,
$$
for $x \in \Lambda_{\hat{\mathbf v}}$, is differentiable and
\begin{align}
\frac{\partial}{\partial \vartheta} \int_{L(x,   \cos(\vartheta) \hv  + \sin(\vartheta) \hv^\perp  )} A \Big |_{\vartheta = 0} = 
\int_{-\infty}^{\infty} \tau B\big ( x + \tau \hv \big )\cdot \big ( \hv^{\perp} \times 
\hv \big ) d\tau +  \Big [ A_\infty (\hv) + A_\infty (- \hv) \Big ]\cdot \hv^{\perp}. 
\end{align} 
\end{lemma}
\noindent{\it Proof:} Take $A^{({\rm SR})} \in \mathcal{A}^{{\rm SR}}_{\Phi}(B) $ and $ \lambda $ such that $A - A^{({\rm SR})} = \nabla \lambda$. Then we have 
\begin{align}\label{lem1}
\int_{L(x,  \cos(\vartheta) \hv  + \sin(\vartheta) \hv^\perp )} A = &  
\lambda_\infty\big ( \cos(\vartheta) \hv  + \sin(\vartheta) \hv^\perp   \big ) 
- \lambda_\infty\big (- (   \cos(\vartheta) \hv  + \sin(\vartheta) \hv^\perp   ) \big ) + \int_{L(x,  \cos(\vartheta) \hv  + \sin(\vartheta) \hv^\perp )} A^{({\rm SR})} \\ \notag  = & \Phi_L(A,  \cos(\vartheta) \hv  + \sin(\vartheta) \hv^\perp ) + \int_{L(x,  \cos(\vartheta) \hv  + \sin(\vartheta) \hv^\perp )} A^{({\rm SR})}. 
\end{align} 
We conclude using Proposition \ref{rem1.} and the fact that 
\begin{align} \label{lem2}
\frac{\partial}{\partial \vartheta}\int_{L(x,  \cos(\vartheta) \hv  + \sin(\vartheta) \hv^\perp )} A^{({\rm SR})} \Big |_{\vartheta = 0}
= \int_{-\infty}^{\infty} \tau B\big ( x + \tau \hv \big )\cdot \big ( \hv^{\perp} \times 
\hv \big ) d\tau,
\end{align} 
which is a direct consequence of Stokes' theorem.   
\bull

\begin{theorem} \label{TPhiA}
Suppose that  $\underline{A} = (V, A)  $, $A \in \mathcal{A}_{\Phi}^{{\rm LR}}(B)$.  The high-momenta limit \eqref{TP1I} of the scattering operator $S(\underline A)$ uniquely determines $  \Phi_L(A, \hv) $ modulo $\pi$, for every $\hv \in \mathbb{S}^2$ ( in the case $A_0 = 0$, it uniquely determines  $  \Phi_L(A, \hv) $  modulo $2 \pi$). If we, furthermore, assume that  $A \in \mathcal{A}_{\Phi,  \delta}(B)  $ (for some $\delta >1$ ), and that $ B \in C^{2}(\overline{\Lambda})$ 
     is such that $ |  B(x)| \leq C \frac{1}{(1 +|x|)^\mu} $, $ | \frac{\partial}{\partial x_i} B(x)| \leq C \frac{1}{(1 +|x|)^{\mu + 1}} $, $ | \frac{\partial}{\partial x_j}\frac{\partial}{\partial x_i} B(x)| \leq C \frac{1}{(1 + |x|)^{\mu + 2}} $, for every 
$i, j \in \{ 1, 2,3 \}$ and every $x \in \Lambda$, then high-momenta limit \eqref{TP1I} of the scattering operator uniquely determines  $ A_\infty(\hv) + A_\infty(-\hv),$  for all $\hv \in \mathbb{S}^2$. 
\end{theorem}
\noindent{\it Proof:}
Take   $ x = R \hv^{\perp}$, where $\left( \kappa^{-1}\big (  \{1\} \big )\right)^c \subset B(0; R)$. 
From the high-momenta limit \eqref{TP1I} of the scattering operator we uniquely determine 
$ e^{- i \int_{- \infty}^{\infty} 2A \cdot \nu (x + \tau \hv) d\tau } $ 
(if $A_0 = 0$, we uniquely determine $  
 e^{- i \int_{- \infty}^{\infty} A \cdot \nu (x + \tau \hv) d \tau}
$). We recall the notation and procedures used in Definition \ref{long-defi}. The result concerning $\Phi_L(A, \hv)$ follows from the fact that
\begin{align} \label{es1}
\int_{- \infty}^{\infty} A \cdot \nu (x + \tau \hv) d\tau =  
\Phi_L(A; \hv)  + \int_{- \infty}^{\infty} A^{({\rm SR})} \cdot \nu (x + \tau \hv) d\tau  =  
\Phi_L(A; \hv) + \lim_{s \to \infty }\int_{ \mathcal{P}(x, \hv, s)} B,
\end{align}
where $ \mathcal{P}(x, \hv, s) $ is a piece of flat plane whose boundary is $ s(c_{(|x|+ 1), \mathfrak{s}}( x, \hv)-x)+x $. As $B$ can be recovered in $ \mathcal{P}(x, \hv, s) $ from the scattering operator, see Theorem 
 \ref{inverse-fields}, then the assertion follows. Now we prove  part of the theorem  that concerns   
 $ A_\infty(\hv) + A_\infty(-\hv)$. It actually follows from Lemma \ref{lem}, since 
\beq \label{A014}
\frac{\partial}{\partial \vartheta} e^{- i \int_{L(x,  \cos(\vartheta) \hv  + \sin(\vartheta) \hv^\perp  )} 2A} \Big |_{\vartheta = 0}  = 2i e^{- i \int_{- \infty}^{\infty} 2A \cdot \nu (x + \tau \hv) d\tau }  \Big [ \int_{-\infty}^{\infty} \tau B\big ( x + \tau \hv \big )\cdot \big ( \hv^{-\perp} \times 
\hv \big ) d\tau +  \Big [ A_\infty (\hv) + A_\infty (- \hv) \Big ]\cdot \hv^{\perp}\Big ],   
\ene  
and we proved above that $  e^{- i \int_{- \infty}^{\infty} 2A \cdot \nu (x + \tau \hv) d\tau }  $ and $  \int_{-\infty}^{\infty} \tau B\big ( x + \tau \hv \big )\cdot \big ( \hv^{\perp} \times 
\hv \big ) d\tau  $ can be recovered from the high-momenta limit
of the scattering operator. Then we recover $   \Big [ A_\infty (\hv) + A_\infty (- \hv) \Big ]\cdot \hv^{\perp} $ for every $\hv^{\perp}$, orthonormal to  $\hv$. As $A_{\infty}(\hv)\cdot \hv = 0$, for every 
$\hv \in \mathbb{S}^2$ (see Proposition \ref{rem1}), we recover $  A_\infty (\hv) + A_\infty (- \hv) $. 
\bull

We now present some definitions and notation, first introduced in \cite{bw}, see also \cite{bw-kg} (we actually use a slight different notation). We do not give all details and motivations of our formalism, see Definitions 7.4, 7.5, 7.9  and 7.10 in \cite{bw} for a full detailed version: Take $R >0$ such that
$K \subset  B(0; R)$. We define the following equivalence relation on
 $\Lambda_{\hv}$: We say that
$ x R_{\hv} y$ if, and only if,
$[c_{\| x \| + \| y \| + R, \mathfrak{s}}(x,\hv)]_{H_1(\Lambda;\ere)}= $ 
$ [c_{  \| x \| + \| y \| + R , \mathfrak{s}'}(y,\hv)]_{H_1(\Lambda;\ere)}$ 
for every curves  $\mathfrak{s}$, $\mathfrak{s}'$, here $ H_1(\Lambda;\ere) $ is the one-singular-homology group in $  \Lambda $ with coefficients in $\mathbb{R}$. Notice that once the equality follows for some  curves $ \mathfrak{s} $, $\mathfrak{s}'$,  it holds true for every such curves, because the sphere $\mathbb S^2$ is simply connected. 
We denote by $\left\{ \Lambda_{\hv, h} \right\}_{h \in \mathcal I}$ the partition of
$\Lambda_{\hv}$ given by this equivalence relation (notice that it is an open disjoint cover of $ \Lambda_{\hv} $). Observe that this equivalence relation, and the associated partition of $\Lambda_{\hv}$ coincide with the one given in \cite{bw}, \cite{bw-kg}.
\begin{definition} \label{def-7.11}{\rm
Let $\Phi$, $ A \in \mathcal{A}^{{\rm LR}}_{\Phi}(0)$, $\hv \in \ese^2$,   and    $ h \in \mathcal I$. Take $R >0$ such that
$ K \subset  B(0; R)$.  We define,
$$
F_h:= \int_{c_{|x| + R, \mathfrak{s}}(x,\hv)} A,
$$
 where $x$ is any point in $\Lambda_{\hv,h}$. Note that $F_h$ is independent of  the    $x \in \Lambda_{\hv,h}$
that we choose, $ R $ and $\mathfrak{s}$. $F_h$ is the flux of the magnetic field over any surface (or chain) in $\ere^3$ whose boundary
is $c(x,\hv)$. We call $F_h$ the magnetic flux on the hole $h$ of $K$.}
\end{definition}

\begin{theorem}\label{th-7.12l} Set  $\phi, \psi \in  {\bf H}_{\langle x \rangle^{4 l}}^2(\mathbb{R}^3)^2$ as in Theorem \ref{TPG}, with $\phi$ compactly supported. Suppose that $A_0 = 0$, $B = 0$.  
For every $A \in \mathcal{A}_\Phi^{{\rm LR}}(0)$    
\begin{align}
\langle U S(\underline A) U^{-1}\, e^{i {\bf x} \cdot v \nu }\phi, e^{i {\bf x} \cdot v \nu }  \psi \rangle = & \sum_{h\in \mathcal I}\,
 \Big \langle   \,  \begin{pmatrix}  e^{i ( F_h + \Phi_L(A, \hv) )}  & 0 \\ 0 &  
  e^{-i( F_h + \Phi_L(A, \hv)  )} \end{pmatrix}  e^{i {\bf x} \cdot v \nu }  \chi_{\Lambda_h} \phi \: ,  e^{i {\bf x} \cdot v \nu } \ \chi_{\Lambda_h} \psi \Big \rangle\\ \notag & + O\left( \frac{1}{v}\right) \|    \phi   \|_{\bf H_{\langle x \rangle^{4 l}}^2 (\mathbb{R}^3)}  
 \|    \psi   \|_{\bf H_{\langle x \rangle^{4 l}}^2 (\mathbb{R}^3)} .
\end{align}
\end{theorem}
\noindent{\it Proof:}
The result follows from Theorem \ref{TPG}, where we  take $A^{(1)} \in \mathcal{A}_\Phi^{{\rm SR}}(0)$ that has compact support (see Remark  3.21 in \cite{bw-kg}) so that the error term in Theorem \ref{TPG} is of order
  $O(\frac{1}{v})$. Furthermore, we take into account that for every $ x \in \Lambda_h : \: $  
$\int_{L(x, \hv)}A =  F_h  - \lim_{s \to \infty }
  \int_{ \mathfrak{s}(s)}  A = F_h +  \Phi_L(A, \hv) $

   \begin{corollary} \label{cor-7.13}
Under the conditions of Theorem \ref{th-7.12l},
the high-momenta limit (6.14) of $S(\underline A)$ in  a single direction $\hv$  uniquely determines
 $\Phi_L(A,\hv)$ and  the fluxes $ F_h, h\in \mathcal I$,  modulo $2\pi$.
\end{corollary}
\noindent {\it Proof:} The Corollary is a direct consequence of Theorem  \ref{th-7.12l}.

\begin{remark}{\rm The results in Theorem  \ref{th-7.12l} are also true in the non-relativistic case of the Schr\"odinger equation in three dimensions considered in \cite{bw}, with a similar proof, and also in the two dimensional case studied in \cite{bw-2dim}, with the necessary changes due to the  differences in the geometry.
}

\end{remark}
\section{Conclusions}
We prove that the high momenta limit of the scattering operator is given by 
\begin{equation}\label{extras}
\begin{pmatrix}
 e^{i \int_{-\infty}^{\infty} dr  (   A \cdot \nu - A_0) ({\bf x} + r \nu ) }  & 0 
\\ 0 & e^{- i \int_{-\infty}^{\infty} dr   ( A \cdot \nu + A_0) ({\bf x} + r \nu ) }
\end{pmatrix}, 
\end{equation}
where ${\bf x}$ is the position operator,  from which we recover the electromagnetic field and  magnetic fluxes modulo $2\pi$. We prove that $A_\infty(\hv) + A_\infty(- \hv)$, for every $\hv \in \mathbb{S}^2$, can be recovered. We, additionally, 
  give a simple formula for the high momenta limit of the scattering operator, assuming the electromagnetic field vanishes (outside the magnet).

The scattering problem that we consider in this paper is important in the context of the Aharonov-
Bohm effect \cite{ab} (see Section \ref{int1}). The issue at stake is what are the fundamental electromagnetic quantities in quantum physics.

In regard to the description of electrodynamics based on non-integrable phase factors 
\cite{wuyang} (see also \cite{Dirac}) the physically significant quantities are gauge invariant and (according also to the experiments of Tonomura) the only observable quantities are the electromagnetic fields and fluxes modulo $2 \pi$.

Our results show that in the relativistic case (the non-relativistic case is studied in \cite{bw-2dim}) the scattering operator contains more information than what can be measured in
experiments. We can uniquely reconstruct from the scattering operator 
$A_\infty(\hv) + A_\infty(- \hv)$, for every $\hv \in \mathbb{S}^2$, which is not invariant by adding to the flux an integer
multiple of $2 \pi$ and it is not either gauge invariant. 

The long range potentials are relevant because a big proportion of the theoretical studies, starting with \cite{ab}, analyse two dimensional models, in which long-range magnetic potentials are unavoidable. Here we deal with three dimensions, but our results are also  valid in two dimensions, with some changes due to the difference in geometry. The two dimensional case was already discussed in \cite{bw-2dim} (in the non-relativistic case). The results on  this  work  show that  our methods in \cite{bw-2dim}   apply to three dimensional models.

\newpage

\begin{figure}\label{fig1}
\begin{center}
\includegraphics[width=17cm]{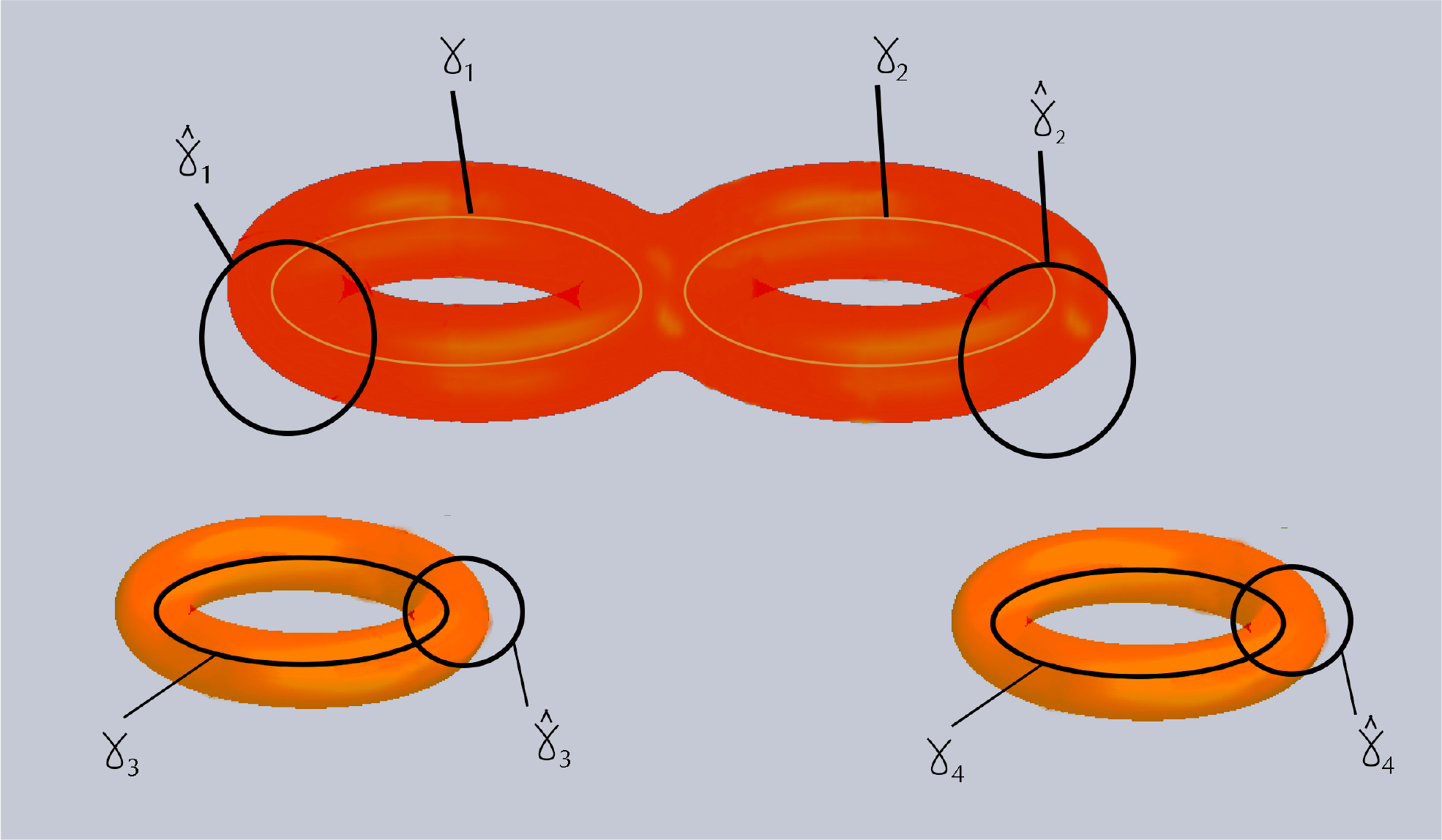}
\caption{The magnet  $ K= \cup_{j=1}^L  K_j \subset \ere^3$  where
$K_j$ are handlebodies , for
every $j \in \{ 1, \cdots, L \}$. The exterior domain,  $\Lambda:= \ere^3
\setminus K$.The curves $\gamma_k, k=1,2,\cdots m$ are a basis of
the first singular homology group of $K$ and the curves
 $\hat{\gamma}_k, k=1,2,\cdots m$ are a basis of the first singular homology group of $\Lambda$. }
\end{center}
\end{figure}
\newpage
\begin{figure}\label{fig2}
\begin{center}
\includegraphics[width=15cm]{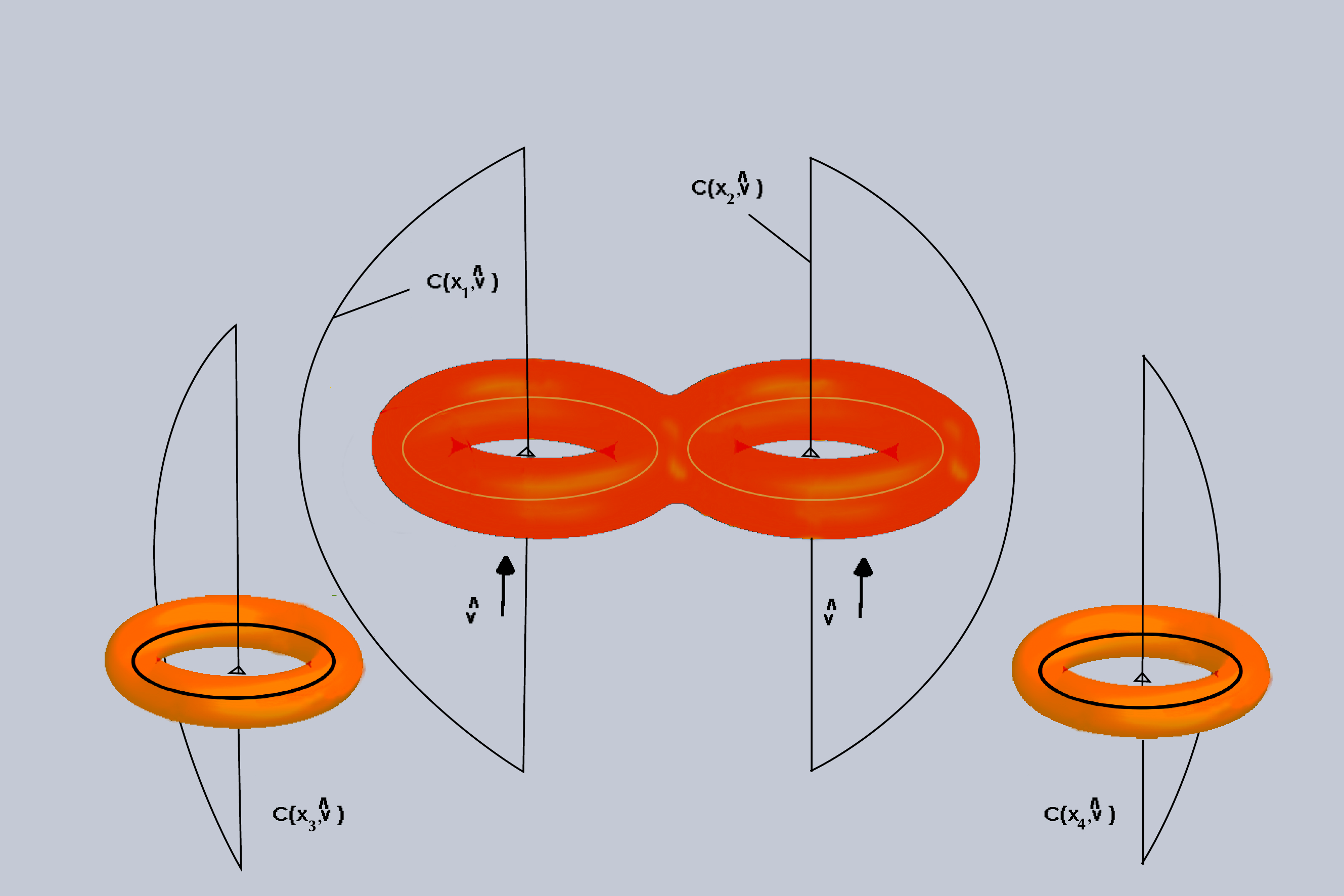}
\caption{The curves $c(x,\hv)$.}
\end{center}
\end{figure}

\end{document}